\pdfoutput=1
\documentclass[natbib]{svjour3}                     % onecolumn (standard format)
\smartqed  % flush right qed marks, e.g. at end of proof
\usepackage{graphicx}
 \usepackage{times}
 \journalname{Space Science Reviews}
\newcommand{\aap}{{Astron. Astrophys.}}
\newcommand{\apj}{{Astrophys. J.}}

\newcommand{\solphys}{{Solar Phys.}}
\newcommand{\ssr}{{Space Sci. Rev.}}
\newcommand{\pasj}{{Publ. Astron. Soc. Japan}}

\newcommand{\apjl}{{Astrophys. J.}}
\newcommand{\nat}{{Nature}}
\newcommand{\mnras}{{Monthly Not. Roy. Astron. Soc.}}

\newcommand{\aapr}{{Astron. Astrophys. Rev.}}
\newcommand{\ao}{{Appl. Opt.}}
\newcommand{\aaps}{{Astron. Astrophys. Suppl. Ser.}}

\newcommand{\zap}{{Z. Astroph.}}
\begin{document}

\title{History of Solar Magnetic  Fields since George Ellery Hale}


\author{J.O. Stenflo}


\institute{Institute of Astronomy, ETH Zurich, CH-8093 Zurich, Switzerland \\ 
              \email{stenflo@astro.phys.ethz.ch}, and \and  \at
Istituto Ricerche Solari Locarno, Via Patocchi, CH-6605 Locarno-Monti, Switzerland}

\date{Received: date / Accepted: date}

\maketitle

%%%%%%%%%%%%%%%%%%%%%%%%%%%%%
\begin{abstract}
As my own work on the Sun's magnetic field started exactly 50 years
ago at Crimea in the USSR, I have been a participant in the field during nearly
half the time span since Hale's 
discovery in 1908 of magnetic fields in sunspots. The present
historical account is accompanied by photos from my personal
slide collection, which show a number of the leading personalities who
advanced the field in different areas: measurement
techniques, from photographic to photoelectric and imaging methods in
spectro-polarimetry; theoretical foundations of MHD and the
origin of cosmic magnetic fields (birth of dynamo theory); the quest
for increased angular resolution from national projects to
international consortia (for instruments both on ground and in space);
introduction of the Hanle effect in astrophysics and the Second Solar
Spectrum as its playground; 
small-scale nature of the field, the fundamental resolution limit,
and transcending it by resolution-independent diagnostics. 
\keywords{Sun: atmosphere \and magnetic fields \and polarization \and dynamo
  \and magnetohydrodynamics (MHD)}
\end{abstract}

%%%%%%%%%%%%%%%%%%%%%%%%%%%%%
\section{Sunspots as a window to cosmic magnetism}\label{sec:sunspots}
Astrophysics can be seen as remote sensing: the information on the
physical conditions in the universe reaches us encoded in the spectra
that we record with our telescopes. Our task is
to decode that information through spectral analysis. 

\begin{figure*}
\centering
\includegraphics[width=0.95\textwidth]{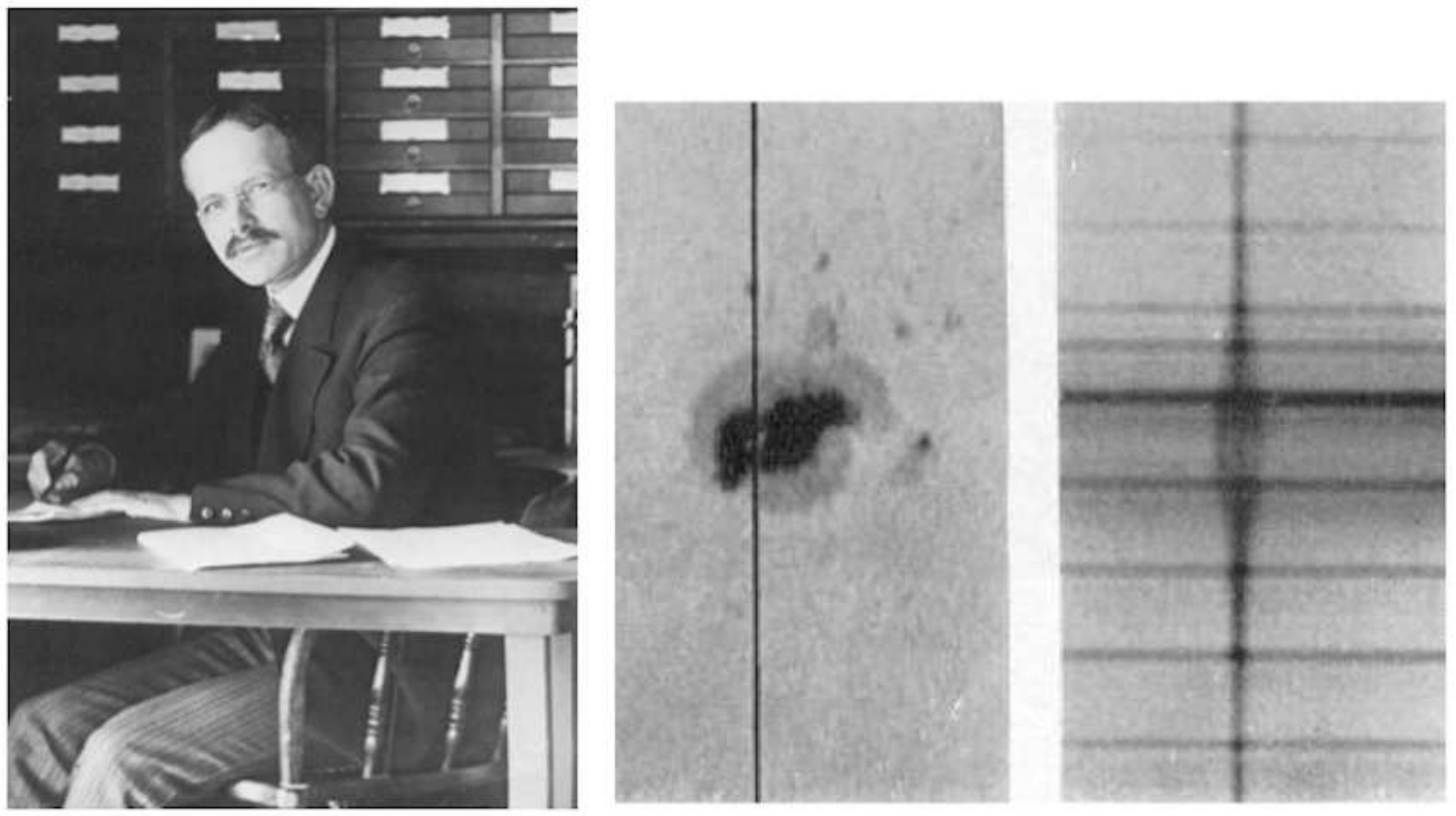}
\caption{George Ellery Hale and his dicovery of magnetic fields in
  sunspots through observations of the Zeeman effect. At the place
  where the spectrograph slit crosses a sunspot the spectral lines get
split in polarized components. The magnitude of the splitting is
proportional to the magnetic field strength, while the polarization
state (circular or linear) reveals the orientation of the field.}
\label{fig:hale} \end{figure*}

With the discovery by Pieter \citet{stenflo-zeeman1897} of the Zeeman effect, it
became known that magnetic fields leave their encoded ``fingerprints''
in the splitting and polarization of spectral lines. Through spectro-polarimetry we
can therefore extract information on magnetic fields in remote objects. George
Ellery Hale was the first to make use of this opportunity in
astrophysics. He had noticed how the shape of the Sun's corona
suggests that the Sun is a magnetized sphere with a global dipole-like
field, and that the filamentary vortex structure of the H$\alpha$
fibrils around sunspots indicate that sunspots are the seats of strong
magnetic fields and vertical electric currents. This led him to look
for and find pronounced Zeeman splitting in sunspots \citep{stenflo-hale08}. 

Sunspots thus became the gateway to the exploration of cosmic magnetic
fields, including the origin of the fields and their role in
astrophysical plasmas. We now know that magnetic fields govern most
cosmic variability on intermediate time scales and pervade all cosmic
plasmas, generating structuring, thermodynamic effects, and
instabilities. Measurements of the Sun's magnetic field has guided the
development of magnetohydrodynamics and dynamo theories. 

The pioneering work by Hale and his team led to the discovery of
Hale's polarity law (the law that governs the E-W orientation of the
sunspot polarities with respect to solar hemisphere and 11-yr cycle),
which showed that the magnetic cycle is 22 yr (the Hale cycle), twice
the length of the sunspot cycle \citep{stenflo-haleetal19}. Another
fundamental discovery in the same paper is 
Joy's law, which tells that the polarity
orientation of bipolar magnetic regions deviates systematically (in a statistical
sense)  from the E-W direction: the orientation is tilted so that the
leading part of the region (with respect to the direction 
of solar rotation) is closer to the equator than the following
part. Hale's and Joy's laws serve as observational
cornerstones of solar dynamo theory. The Sun's dynamo can be seen as
a prototype for all cosmic dynamos.

%%%%%%%%%%%%%%%%%%%%%%%%%%%%%
\section{The enigmatic general magnetic field of the Sun}\label{sec:general}
Soon after his discovery of magnetic fields in sunspots Hale wanted to
explore the non-spot ``background field'', the so-called ``general''
magnetic field of the Sun, which he believed to be a global dipole
based on the appearance of the solar corona at eclipses. The
observational technique
was to place a grid of mica strips across the spectrograph slit, so
that one gets an alternating sequence of left- and right-handed
circularly polarized spectra. In the presence of a magnetic field
there would be a relative line displacement between the opposite polarization
states. 

The spectra were recorded on photographic plates, and the line
positions were determined manually with a micrometer. With modern
standards this was a crude technique not adequate for detection of the weak
background fields. Aware of the marginal nature of this undertaking
and the danger of subjectivity in the results, Hale
let many of his assistants independently reduce the same plate
material, and he claimed in the publications that the persons reducing
the data were not informed about which heliographic latitude a given
plate referred to. This was supposed to eliminate possible bias from the
results. 

%%%%%%%%%%%%%%%%%%%%%%%%%%%%%
\subsection{An example of personal bias}\label{sec:bias}
All of the assistants except two got null results (consistent with
what is to be expected in view of the measurement uncertainties). One
of these two was a young postdoc who had recently arrived to Pasadena
from Holland, Van Maanen. He immediately impressed Hale by getting
positive results, and Hale gave most weight to his findings. This resulted
in claims that the strength of the polar field was about an order of
magnitude stronger than we find with modern, far 
more accurate techniques \citep{stenflo-haleetal18}. In addition the
polarity of the polar fields 
was opposite to what one would expect according the much later
discovered model of the solar cycle by Babcock (1961). 

In 1968, shortly after completion of my Doctoral Thesis with the title
``The Sun's Magnetic Field'', I spent seven months in Pasadena with the Mt
Wilson Observatory and decided to use this opportunity to explore the
Hale archives in search for an explanation of his enigmatic results on the
Sun's general magnetic field. I was able to locate 400 photographic
plates recorded in 1914, together with the meticulous notebook of Van
Maanen with his micrometer measurements of the Zeeman shifts for each
of these plates. After having the plates carefully cleaned I took them
in my car and drove to the Sacramento Peak Observatory (Sunspot, New
Mexico), where they had a digitized microphotometer with which I could
(with the help of Jacques Beckers) have all the plates scanned. I then
brought the big magnetic tapes with the scanned data to Lund, Sweden,
where I analyzed them with a CDC mainframe computer. My computer
reduction gave null results (within the standard error) for the
general magnetic field, while Van Maanen had found for the same set of
plates a field much stronger than the standard error, which varied
with heliographic latitude as a global dipole field with the wrong 
polarity \citep{stenflo-s70hale}. 

Around the time of Hale's pioneering solar work the astronomical
community was involved in an intense debate on the nature of spiral
nebulae: Were they ``island universes'' (what we now call galaxies), or
were they nearby objects inside our Milky Way system\,? Van Maanen
joined this debate and provided what at that time seemed to be very
convincing evidence against the island universe idea. Around 1922-1923
he measured on photographic plates substantial proper
motions along the spiral arms of M33, which implied that the rotation
period of M33 was smaller than that of the Milky Way by a factor of
one thousand and that M33 could therefore not be very distant and large
but had to be inside
our galaxy \citep[cf.][]{stenflo-lundmark26}. We now know that M33 is a galaxy and that
whatever Van Maanen measured was not a real property.

%%%%%%%%%%%%%%%%%%%%%%%%%%%%%
\subsection{The photoelectric magnetograph}\label{sec:photoelectric}
The general, background magnetic fields, which we now refer to as
quiet-sun magnetic fields, gave too weak Zeeman-effect signatures to
be measurable with the technology that was available to Hale. Further
progress had to wait until the breakthrough in sensitivity provided by
the photoelectric magnetograph, introduced by
\citet{stenflo-babcock53}. Instead of manually measuring wavelength
positions 
of spectral lines on photographic plates one used photomultipliers as
detectors and electro-optical polarization modulation with an ADP-type
Pockels cell. Two exit windows in the spectral focal plane selected
the light from the blue and red wing of a chosen Zeeman-sensitive
spectral line. A servo-controlled tilting glass plate near the
spectral focal plane kept the spectral line centered with respect to
the two exit slits. The tilt angle of the plate represented the
measured Doppler shift. The modulated polarization signals from the
two exit slits were combined in a difference amplifier (since the
polarization signals from the two line wings have opposite signs ---
today we say that the Stokes $V$ profile is anti-symmetric) and then demodulated. 

Since photomultipliers are ``1-pixel detectors'', a
circular-polarization image (which represents a ``magnetogram'') could
only be built up through raster scanning of the solar image. The
initial observations were crude, with a spatial resolution of order
one arcmin, but they allowed the recording of full-disk magnetograms
with a polarimetric sensitivity that was sufficient for mapping the
weak background fields. Thus Babcock could initiate a synoptic
full-disk magnetogram program at the Mt Wilson Observatory. 

This synoptic program led Babcock to the discovery that the general
magnetic field, in particular the field in the polar caps of the Sun,
reverses sign every 11 years. Thus not only the sunspot polarities but
also the global magnetic field varies with the 22-yr Hale cycle. The
photoelectric magnetograph gave us a comprehensive empirical picture
of solar magnetism. The underlying mechanism that governs the Hale
cycle must physically link the behavior of sunspots with the behavior
of the global field. 

\citet{stenflo-babcock61} could integrate the various empirical results into a
phenomenological theory for the solar cycle, which explicitly clarified
how the bits and pieces were connected. The Hale cycle represents an
oscillation between a poloidal and a toroidal configuration of the magnetic
field. The frozen-in field lines of the poloidal field are wound up by
the latitudinal shear of the Sun's differential rotation, so that a
toroidal field is built up. When the strength of the amplified
toroidal field exceeds a certain theshold, magnetic tension can no
longer prevent buoyant sections of the subsurface flux ropes to float up
to the surface to form bipolar magnetic regions with polarity
orientations in the general E-W direction, with opposite orientations
in the N and S hemispheres, as required by Hale's polarity law. During
the rise to the surface the flux loop is rotated by the Coriolis
force, resulting in a systematic tilt of the bipolar orientation with
respect to the E-W direction, in qualitative agreement with Joy's
law. The tilt is such that the poleward (following) part of the bipolar region has
a polarity that is opposite to that of the polar fields in the same
hemisphere. When bipolar regions age, they fragment and spread by
turbulent diffusion to form the general background field. Because of
the initial tilt of the bipolar regions it is the following polarity
that will dominate the new background field at high latitudes, thereby
cancelling and reversing the polar field. Coriolis forces acting on
buoyant toroidal flux combined with turbulent diffusion thus lead to
a regeneration of a poloidal field with reversed polarity. 

While Babcock's model was phenomenological, qualitative, and
descriptive, it contained the physical ingredients that are
central to modern dynamo theories. \citet{stenflo-leighton69} significantly
extended Babcock's model by giving it a formulation in terms of a set
of semi-empirical equations that could be used for quantitative
modeling of the evolution of the pattern of surface magnetic
fields. The Babcock-Leighton framework has been used since the early
1980s by Neil Sheeley and coworkers at NRL \citep[e.g.][]{stenflo-sheeley85}
for systematic modeling of the field pattern to understand its role
for coronal holes, solar wind streams, and other large-scale phenomena
on the Sun.

%%%%%%%%%%%%%%%%%%%%%%%%%%%%%
\subsection{The magnetic network}\label{sec:network}
With the superior sensitivity of the photoelectric magnetograph one
might think that the days of doing magnetometry with photographic
plates were over. However, photographic plates represented 2-D
detectors with an enormous number of ``pixels'', which for mapping
purposes with high angular resolution represented a huge advantage
over photomultipliers, which are 1-pixel devices. 

\begin{figure*}
\centering
\includegraphics[width=0.95\textwidth]{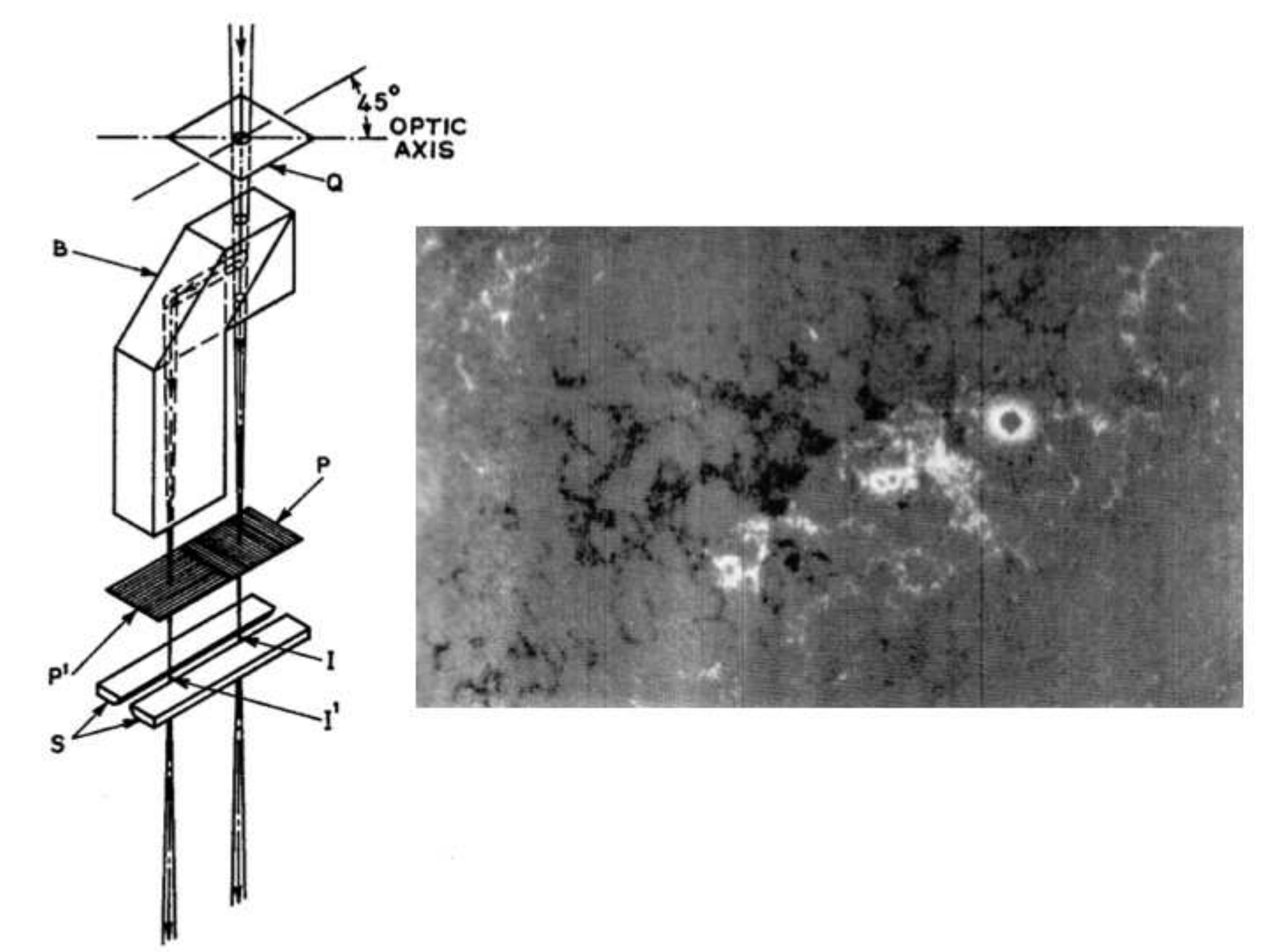}
\caption{The principles of the Leighton photographic magnetograph are
  illustrated to the left, from \citet{stenflo-leighton59}. With a
  polarizing beam splitter two images 
in opposite circular polarization states of the chosen solar region
are formed at the entrance slit of the spectrograph. The photographic
plate is located behind the exit slit in the spectral focus. The
entrance and exit slits are scanned in unison to build up
monochromatic images (spectroheliograms) on the photographic
plate. After the plate has been developed, the two
orthogonal-polarization images are subtracted, to give a
magnetogram like the one in the right part of the figure
\citep[from][]{stenflo-vrabec71}. It reveals the network structure of
the magnetic-field pattern.} 
\label{fig:leighton} \end{figure*}

Robert Leighton developed a photographic magnetograph technique
based on the photographic recording of two simultaneous spectroheliograms in
orthogonal circular polarization states \citep{stenflo-leighton59}, as
illustrated in the left portion of
Fig.~\ref{fig:leighton}. Subsequent photographic 
subtraction of the two images then results in a polarization map
(line-of-sight magnetogram). Such photographic subtraction is of
course a tricky business due to the non-linear response of
photographic plates, but Leighton and his followers developed it into
an art. 

One place in particular where the Leighton technique was perfected to produce not
only high-resolu\-tion maps but also magnetic-field movies of
outstanding quality was the San Fernando Observatory. The right part
of Fig.~\ref{fig:leighton} gives an example, from \citet{stenflo-vrabec71}, of
such a magnetogram, which clearly reveals that the magnetic pattern has a
network structure. Remarkable movies of the
moat structure around sunspots were made, showing how flux fragments of opposite
polarities stream away from the spots. 

Leighton also used his subtractive spectroheliogram technique to make
full-disk Doppler maps of the Sun, which led to his discovery of the
supergranulation \citep{stenflo-leightonetal62}. The largely
horizontal flows in the supergranulation 
were much more effective than the granular flows for the
redistribution of the fields through ``turbulent diffusion''. They came
to play a major role in Leighton's kinematic model of the solar cycle,
in which he treated the redistribution as a random walk on
the solar surface of the frozen-in field lines. 

\begin{figure*}
\centering
\includegraphics[width=0.77\textwidth]{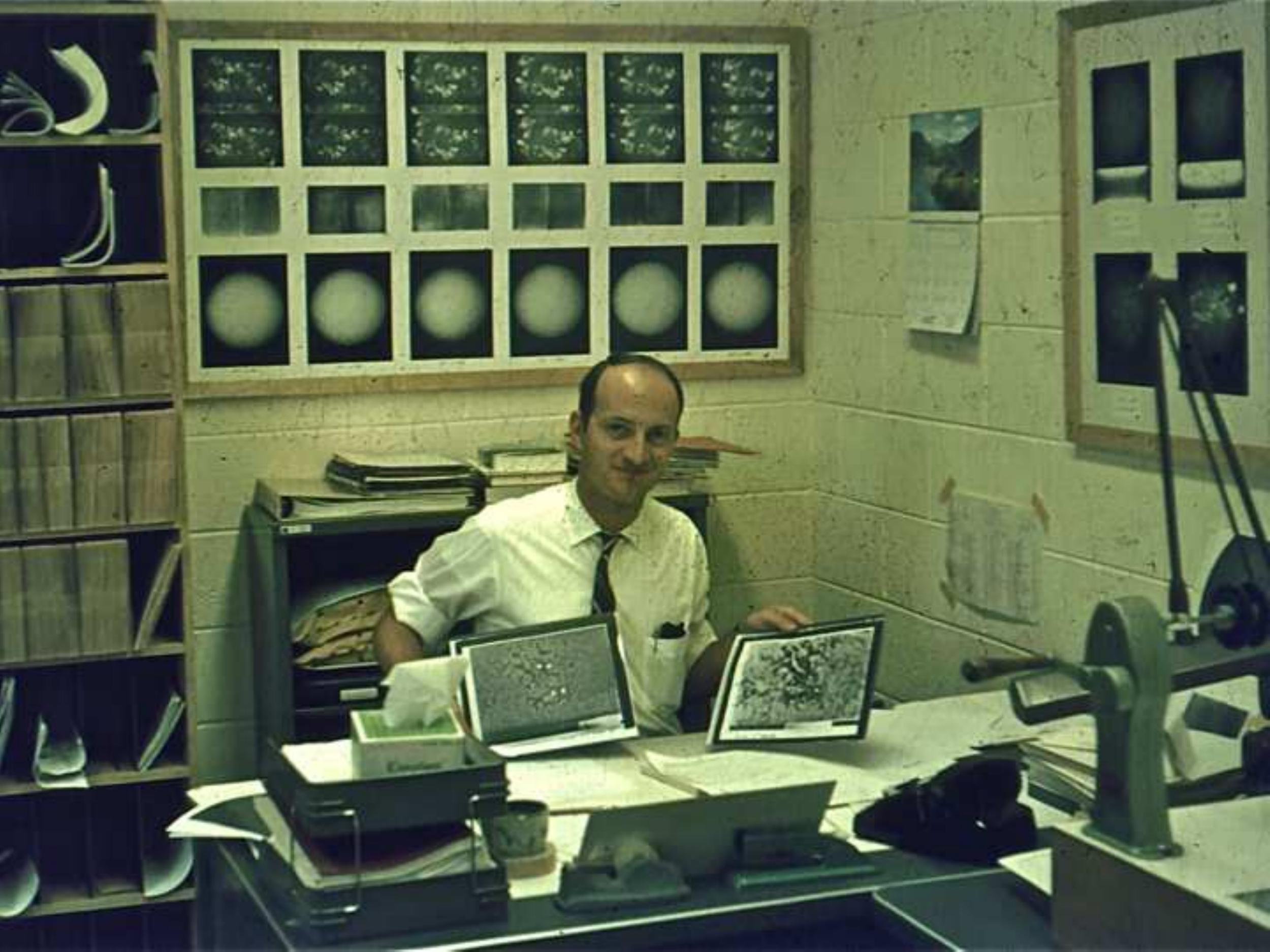}
\caption{Neil Sheeley in his Tucson office in 1968. He perfected the
  art of recording Kitt Peak spectroheliograms to explore the photospheric
  network. }
\label{fig:sheeley} \end{figure*}

Spectroheliograms had since many decades been an established tool to make
narrow-band monochromatic images of the Sun for any chosen wavelength
in the visible solar spectrum. They had revealed that chromospheric
lines like the H and K lines of ionized calcium show a network-like
emission pattern. Leighton's Doppler and magnetic-field maps showed
that both the supergranular velocity network and the magnetic-field
network were closely correlated with the emission network, they seemed
to coincide with each other. The statistical correlations between these three
types of network were intensely explored by many solar physicists around
the end of the 1960s, in particular by recoding Doppler and magnetic
data through raster scans with a photoelectric magnetograph, which
generally allowed the analysis to be more quantitative than with the
photographic method. 

In 1967 Neil Sheeley published a seminal paper in which he showed that
many photospheric lines had line gaps (very significant line
weakenings) at the locations of the quiet-sun network, and that these
were places with strong, small-scale flux concentrations, with field
strengths reaching up to hundreds of G 
\citep{stenflo-sh67}. In spectroheliograms the 
line gaps show up as a photospheric emission network. Subsequently
\citet{stenflo-chsh68} made an extensive spectroheliogram survey of
the behavior of different photospheric spectral lines with various
excitation potentials, to demonstrate that the line gaps were the
result of increased photospheric temperature. The observations allowed the
temperature structure of the photospheric network to be modeled. 

Sheeley's hundred G quiet-sun field strengths were shockingly high for
that time, but they still represented a lower limit determined by the
spatial resolution of his instrument. Since the intrinsic sizes of the
network elements were not known and therefore not their filling
factors $f$, the intrinsic field strength would be $1/f$ times the
apparent field strength. 

To find the intrinsic field strength without achieving infinite
angular resolution one had to come up with a resolution-independent
approach, which occurred soon afterwards with the introduction of the
magnetic line-ratio method (cf. Sect.~\ref{sec:mcmath}). 

\begin{figure*}\sidecaption
\includegraphics[width=0.43\textwidth]{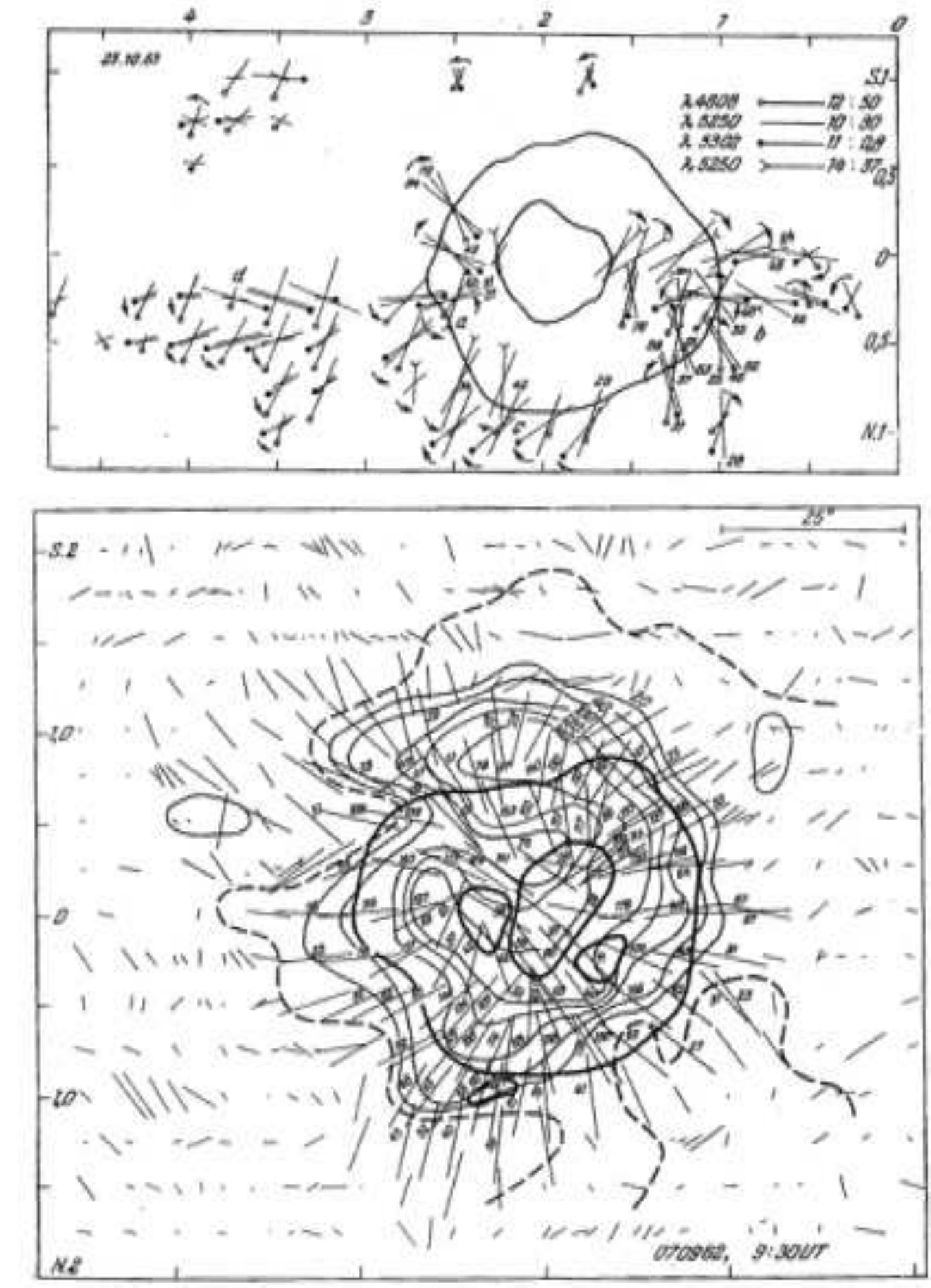}
\caption{Examples of maps of vector magnetic fields in regions around
  sunspots, from \citet{stenflo-severny66}. The top panel shows how the field
  azimuth rotates as observed with four spectral lines formed at
  different heights. The contours represent the sunspot umbra and
  penumbra. In the bottom panel the thin solid contours represent 
  the longitudinal magnetic field, superposed on the background of
  short lines that represent the strength and orientation of the
  transverse magnetic field.}
\label{fig:vectormap} \end{figure*}

%%%%%%%%%%%%%%%%%%%%%%%%%%%%%
\section{Mapping the vector magnetic fields and electric
  currents}\label{sec:vector} 
Already Hale in his early work on sunspot magnetic fields knew how to
combine the measured circular and linear polarizations of the 
longitudinal and transverse Zeeman effects to derive the magnetic-field vector
(strength and orientation of the field). Formally such a Zeeman-effect
``inversion'' is very straightforward. In practice the inversion often leads
to false results because of a combination of factors, in particular non-linear response,
non-Gaussian noise, and unknown subresolution structure of the
field. In contrast to the transverse Zeeman effect, the polarization
of the longitudinal Zeeman effect has a nearly linear dependence on
the field (via its line-of-sight component) with a Gaussian noise
distribution, and also has high field sensitivity. This is the reason
why most observational explorations of the magnetic field has been of its
line-of-sight component. What we call ``magnetogram'' is always a 
circular-polarization map; in the more rare cases when also the Zeeman effect
in the linear polarization is being recorded we use the term ``vector
magnetogram''. 

Soon after Babcock's introduction of the photoelectric magnetograph
his concept was being extended at various places, including Crimea,
IZMIRAN (near Moscow), Pulkovo by Leningrad, and Ond\v{r}ejov by Prague,
to record vector magnetic fields. Both the circular and linear
polarizations in a selected spectral line were modulated by Pockels
cells. The demodulated signals were recorded with ink pen on
strip-chart recorders and combined to 
give the magnetic-field vector. The pioneering work in vector
magnetometry was done at the Crimean Astrophysical Observatory, where
\citet{stenflo-stepseverny62} published the first vector-field maps of
the Sun. Figure \ref{fig:vectormap} from \citet{stenflo-severny66} shows an
example of maps of the vector magnetic field around sunspots 
\citep[see also][]{stenflo-severny64}. 

Vector magnetometry was a hot topic in solar physics of
the 1960s, because it was seen as a tool for the exploration of the
full MHD structure of the solar atmosphere. Maxwell's equations were
being used to derive maps of the vertical electric current density
from the horizontal gradients of the observed transverse magnetic
fields. The pattern of vertical electric currents was found to correlate 
with the locations where flares occurred, as expected in the current-interruption
theory of solar flares by \citet{stenflo-alfvencarlqvist67}. Around that
time Alfv\'en was finding it more revealing to describe MHD
environments in terms of the electric current structure instead
of magnetic fields, since currents are responsible for
the pinch effect and could cause instabilities with the development
of double layers, in which the MHD approximations break down. 

The upper panel of Fig.~\ref{fig:vectormap} indicates how the apparent
transverse field rotates with height in the atmosphere. Superposing the
results obtained with four spectral lines formed at different heights
one reveals how the azimuth angle varies. This behavior suggests that the
field lines are twisted, as expected in the presence of vertical
electric currents. However, the interpretation has to also account for
the apparent twist induced by magneto-optical effects.  

\begin{figure*}\sidecaption
\includegraphics[width=0.45\textwidth]{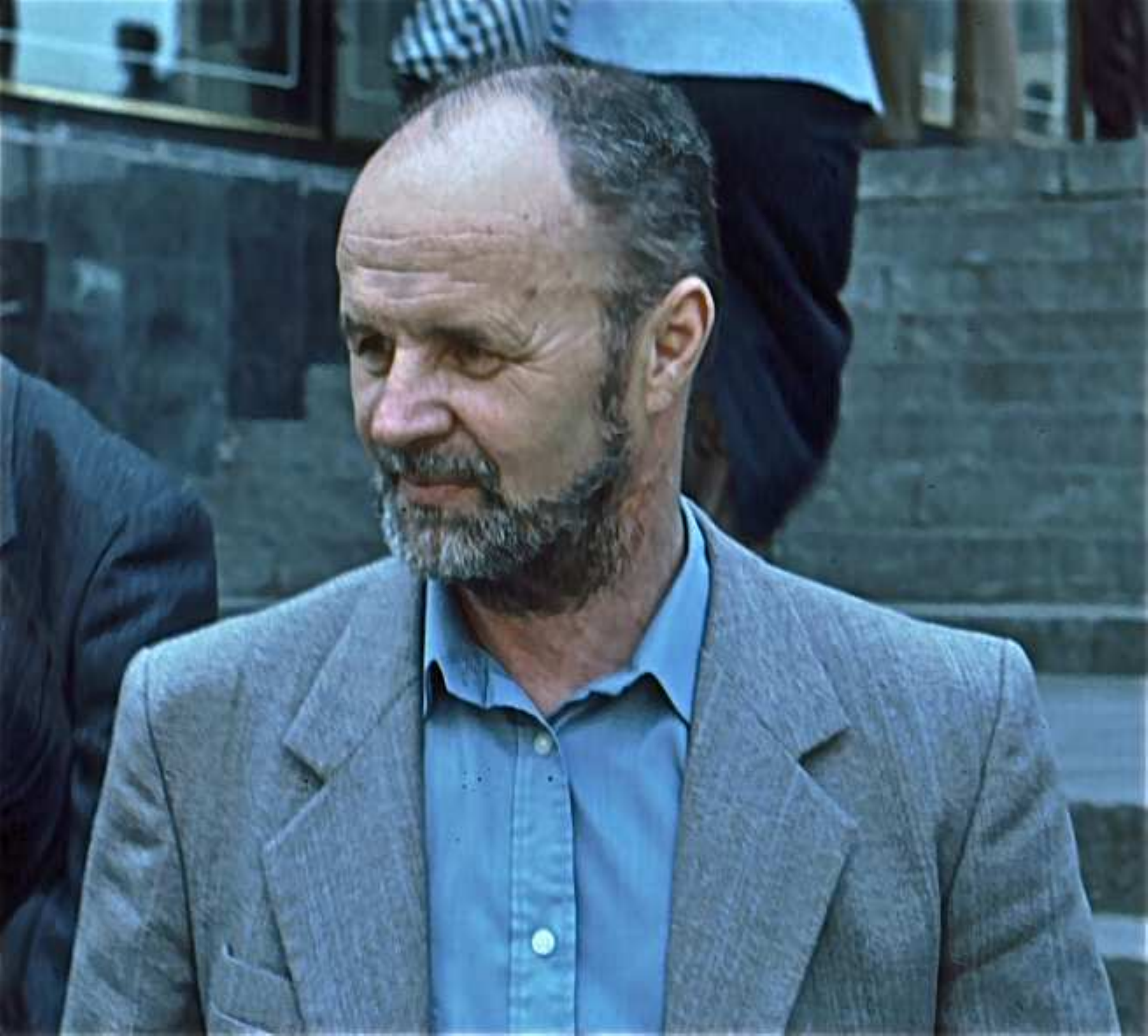}
\caption{Dimitry Nikolaevich Rachkovsky (1928-2013), who gave us a
  comprehensive formulation of LTE radiative transfer of the full
  Stokes vector in magnetized 
  media, including magneto-optical 
  effects. His formulation is used in all current Stokes inversion
  work. The photo was taken in 1989 in Kiev.}
\label{fig:rachkovsky} \end{figure*}

The Crimean Astrophysical Observatory was not only the world-leading
institute in vector magnetometry, it was also the place where the
theoretical foundations for the interpretation of magnetograph
measurements were developed. While Wasaburo Unno in Japan had
published a phenomenological but correct derivation of the radiative
transfer equation for the Stokes vector \citep{stenflo-unno56}, it was 
the work of Stepanov and Rachkovsky at Crimea that brought us a 
self-consistent mathematical derivation that also included the magneto-optical
effects. While Stepanov formulated this approach
%\citep{stenflo-stepanov58a,stenflo-stepanov58b}, 
(Stepanov 1958a,b), 
Rachkovsky completed the theory to give us a mathematically stringent and
comprehensive (within the assumption of LTE) formulation of the Stokes
vector radiative transfer problem
%\citep{stenflo-rachkovsky62a,stenflo-rachkovsky62b}. 
(Rachkovsky 1962a,b). 
Rachkovsky's theory was used to 
identify observational signatures of magneto-optical effects in the
recordings with the Crimean magnetograph. 

\begin{figure*}\sidecaption
\includegraphics[width=0.45\textwidth]{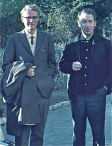}
\caption{At the Crimean Astrophysical Observatory in 1965: Jan Stenflo
and the observatory director, Andrei Borisovich Severny, the leading
Soviet solar physicist at the time.}
\label{fig:severny} \end{figure*}

In 1965 the academies of sciences in Sweden and the Soviet Union
signed an exchange agreement. It happened, thanks to the contacts and
support of my mentor Hannes Alfv\'en, that I was the first from the
Swedish side to be sent over to work in the USSR under this
agreement. Alfv\'en believed that the Sun's magnetic field had a
filamentary fine structure induced by the pinch effect of electric
currents, and that this fine structure invalidated Babcock's
observations since it was not resolved. He conjectured that if the
field consisted of ``microspots'' with less visibility because of
their different thermodynamic properties, then the spatial averaging
process would lead to the wrong interpretation of the
measurements \citep{stenflo-alfven67}. 

Alfv\'en had a strong incentive to see Babcock's results
invalidated, since a general magnetic field that reversed sign every
11 years contradicted Alfv\'en's own theory of the solar cycle. He
sent me to Crimea with the mission to confirm his conjecture by
exploring the fine structure with the Crimean magnetograph. My mission
failed, since my work rather helped to validate Babcock's results. However,
Alfv\'en's ideas put me on a path to explore the nature of the unresolved
magnetic fine structure and how it can be properly diagnosed. This
path took me in productive directions that none of us could foresee at
that time. 

Thus at age 22 I spent three summer months in 1965 to study the fine
structure of quiet-sun magnetic fields with the Crimean magnetograph,
and I returned for another four months in the summer of 1966. This
work gave the observational basis for my Doctoral Thesis, 
which I completed in the spring of 1968 (at the University of Lund,
Sweden). Figure \ref{fig:severny} shows me during the visit in 1965
together with the Director of the Crimean Astrophysical Observatory,
Andrei Borisovich Severny, who was the most prominent Soviet solar
physicist at the time. 

\begin{figure*}
\includegraphics[width=0.95\textwidth]{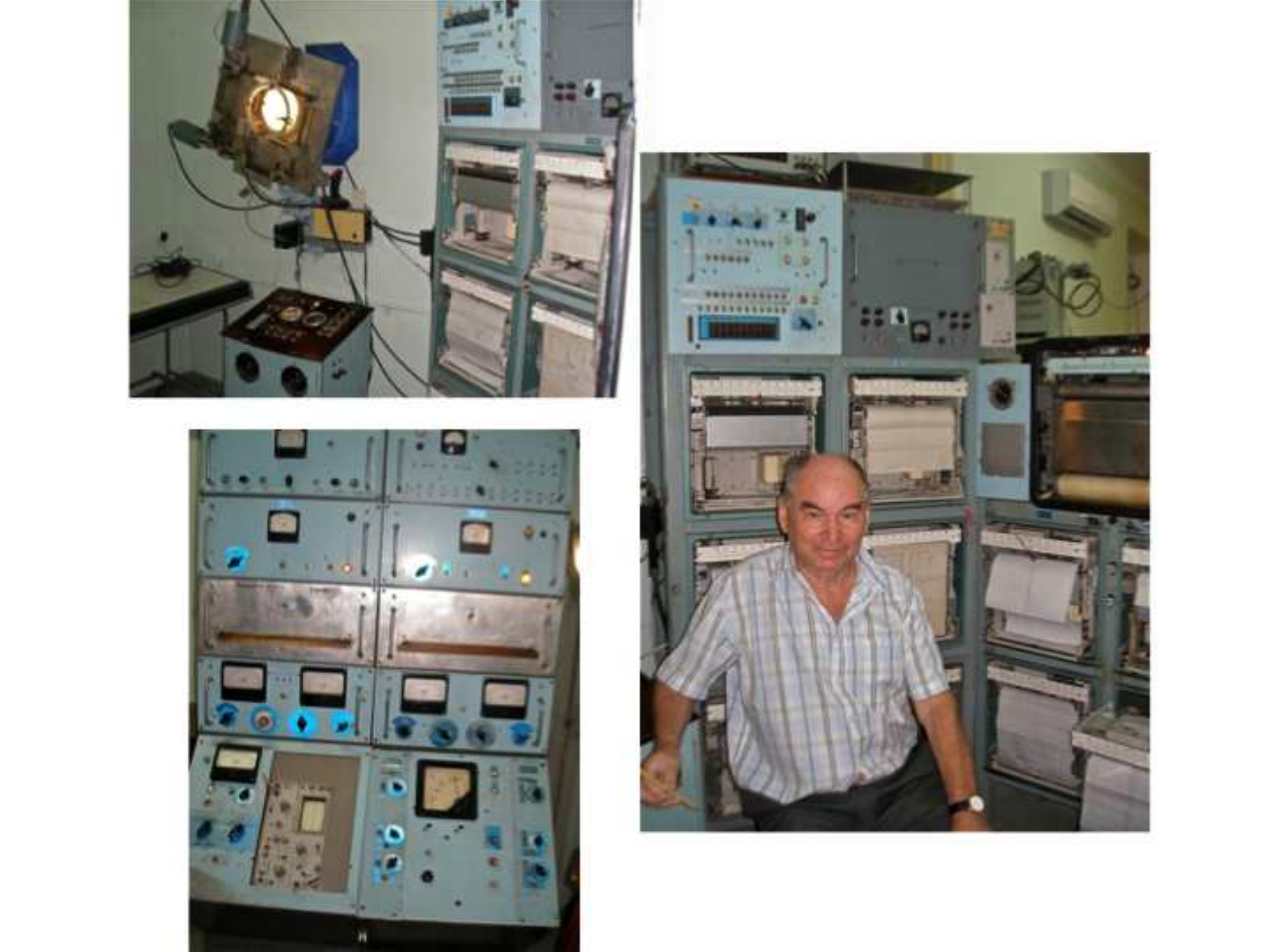}
\caption{Control panels and strip-chart recorders of the Crimean
  magnetograph system in photos taken in 2011. The equipment has not
  changed much since the mid 1960s. In the right panel is Valery
  Kotov, who used the instrumentation in the 1960s to determine the
  vector electric current structure of active regions, and who is now
  using it for continued studies of the Sun's 160-min oscillations.}
\label{fig:kotov} \end{figure*}

The Crimean Astrophysical Observatory was a world leader in solar
physics in the 1960s, it was an exciting time to be there. However,
towards the end of the decade began a period of political and economic
stagnation and decline, which finally ended in the collapse of the
Soviet Union. Figure \ref{fig:kotov} contains photos that I took of the
solar magnetograph when I visited Crimea again in 2011. The electronic
equipment, control panels, and strip-chart recorders were much the
same as when I worked there 45 years before. It was like going back in
time, not much had changed. In the right panel is Valery Kotov, who
still uses this equipment to study the 160-min oscillations of the
Sun. I became befriended with him in 1965, when he was a PhD student
at Crimea. In his thesis on the electromagnetic structure of active
regions he used vector magnetic-field mapping not only to determine
the vertical electric-current density, but also the horizontal current
density and therefore the {\it vector electric currents}. The
horizontal electric currents were found by recording the transverse
magnetic field at different heights, and then from the height
gradients of the horizontal field components use Maxwell's equations
to obtain the horizontal components of the current. 

Since there has been a revival in the mapping of vector magnetic
fields in recent years in the context of Stokes inversions, one may
wonder how reliable all the pioneering results from the 1960s about
the vector magnetic fields were. My general answer is that the maps of
the relative distributions of azimuth orientations of the transverse
fields are reliable (as long as one accounts for the magneto-optical
effects in the interpretation). Likewise the maps of the line-of-sight flux
density (regular magnetograms) are also reliable. However, the
magnitudes of the transverse flux density have been greatly
overestimated, with the result that the determined field inclinations are much
too large, the fields appear much more transverse than they
really are. 

The main source of the problem is that the relation between the
measured linear polarization and the transverse field strength is
highly non-linear (in contrast to the case for the circular
polarization). This has two consequences for observations that do not
have high S/N ratio: (1) The spatial
averaging over unresolved subpixel structures depends on the nature of
this structuring. Ignoring this problem always leads to errors in the
direction  of artificially enhancing the transverse fields. (2) The noise
distribution of the transverse fields is highly non-gaussian, with an
extended non-gaussian tail that makes the fields look much more
transverse than they are. Since in addition the polarization
amplitudes that are produced by a given field strength are much
smaller (by typically a factor of 25) for the linear polarization
relative to the circular polarization, the noise infiltrates the transverse
fields much more than the longitudinal fields, always in the direction
of making the fields appear much too transverse. 

This problem does not go away by using improved Stokes inversion
techniques, because it is not a technical problem. The only known way
to fully eliminate the transverse-field bias (in the weak-field
quiet-sun case) is to use a statistical
approach with ensemble averages (cf. Sect.~\ref{sec:ensemble}),
averaging the Stokes $Q$ linear polarization 
as observed away from disk center (to break the symmetry), without
trying to convert $Q$ to transverse field strength, since such 
conversion is the source of the non-linearities and the non-gaussian
noise.  Because the ensemble averages 
are independent of the spatial resolution of the instrument (since $Q$
is proportional to the number of photons received), and because the
noise distribution in $Q$ is Gaussian and thus symmetric, the ensemble
averages of Stokes $Q$ give unbiased results that neither favors 
horizontal nor vertical. Application of this technique has shown that
quiet-sun magnetic fields are preferentially vertical (with respect to
the isotropic angular distribution) in the low to middle photosphere
(as observed in photospheric lines over most over the solar disk) but
become preferentially horizontal in the upper photosphere \citep[as observed
close to the solar limb, cf.][]{stenflo-s13aa1}.

%%%%%%%%%%%%%%%%%%%%%%%%%%%%%
\section{Origin of the Sun's magnetic cycle}\label{sec:origin}
We have seen how Babcock integrated the various empirical findings
into a phenomenological model of the solar cycle, which contained the
main physical ingredients that govern the
activity cycle of the Sun in terms of an interplay between turbulence and
magnetic fields in a rotating magnetized plasma. We are now convinced that
these ingredients are responsible for the generation of large-scale
magnetic fields all over the universe, not only in stars and planets,
but also on galactic scales. The Sun's 22-yr magnetic cycle is a
prototype for all cosmic dynamos. 

\begin{figure*}
\centering
\includegraphics[width=0.8\textwidth]{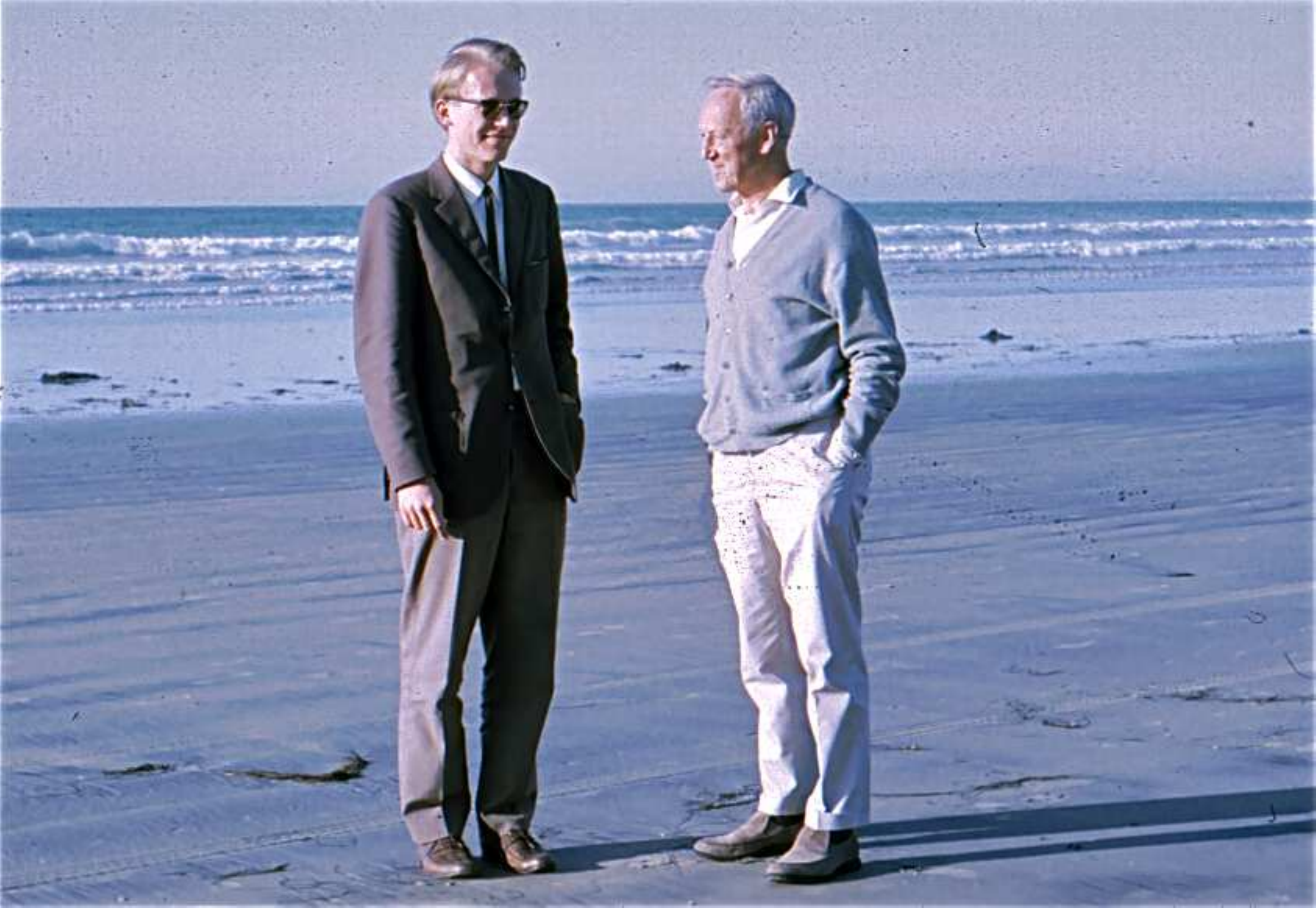}
\caption{Jan Stenflo with his mentor Hannes Alfv\'en on the Pacific
  beach in La Jolla in 1968. Two years later Alfv\'en received the
  Nobel Prize in Physics for his discovery of what we now refer to as
  Alfv\'en waves.}
\label{fig:alfven} \end{figure*}

%%%%%%%%%%%%%%%%%%%%%%%%%%%%%
\subsection{An incorrect theory worthy of a Nobel
  Prize}\label{sec:nobel}
In a Letter to Nature in 1942 (my birth year) my mentor Hannes
Alfv\'en presented (two decades before Babcock) an alternative
explanation for the origin of the sunspot cycle 
\citep{stenflo-alfven42}. In his 
scenario the Sun possessed a global dipole-like field that was a 
remnant from the formation of the solar system. This field therefore
penetrated the core of the Sun, so it could not reverse (in
contradiction to Babcock's later findings). At that time it was
believed that the CNO cycle rather than the p-p reaction chain was
responsible for the thermonuclear energy production. This had the
consequence that the central temperature was considered to be as high
as 25 million K, and that a steep temperature gradient would cause the
solar core to be in a convective state. Alfv\'en therefore postulated
that the sunspots originated from disturbances in the central regions
of the Sun, and that these disturbances traveled as waves (later called
Alfv\'en waves) along the global dipole field, emerging at the surface
as sunspots. 

\begin{figure*}\sidecaption
\includegraphics[width=0.55\textwidth]{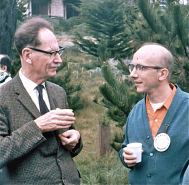}
\caption{T.G. Cowling from Leeds, UK, in conversation with R.F. Howard
  from Mt Wilson Observatory, Pasadena, during the First Solar Wind
  Conference at Asilomar, California, in 1971. Cowling is best known
  for his anti-dynamo theorem \citep{stenflo-cowling33}. Bob Howard
  took over and modernized 
  Babcock's work on recording full-disk solar magnetograms and
  developed the successful synoptic program of Mt Wilson
  magnetograms that has for many decades provided fundamental data on the
  global evolution of the magnetic field.}
\label{fig:cowling} \end{figure*}

We now know this explanation of the solar cycle to be
incorrect. However, in this short paper Alfv\'en for the first time
introduced the concept of MHD waves, explaining why disturbances in a
magnetized plasma would propagate as transverse waves along the field
lines with a speed that we now refer to as the Alfv\'en velocity. It
was for the discovery of these waves that he received the 1970 Nobel Prize
in Physcis. Although he introduced this wave concept within the
context of an incorrect theory for the solar cycle, it was a
Nobel-prize worthy discovery that has since played a fundamental role
in all of plasma physics. It is an example of how wrong
theories can contain elements of great value\,!

%%%%%%%%%%%%%%%%%%%%%%%%%%%%%
\subsection{Beginnings of dynamo theory}\label{sec:dynamo}
The concept of explaining the origin of macroscopic magnetic fields in
terms of dynamo processes goes back to \citet{stenflo-larmor1919}, but
subsequent progress towards a viable theory was relatively slow. One serious obstacle
was the ``anti-dynamo theorem'' of \citet{stenflo-cowling33} (cf. Fig.~\ref{fig:cowling}), 
which was a proof that an axially symmetric velocity field, like that of a
rotating medium, was incapable of generating a dynamo. Such idealized
geometries were insufficient, the symmetry had to be broken for dynamo
action to occur. 

\begin{figure*}
\centering
\includegraphics[width=0.85\textwidth]{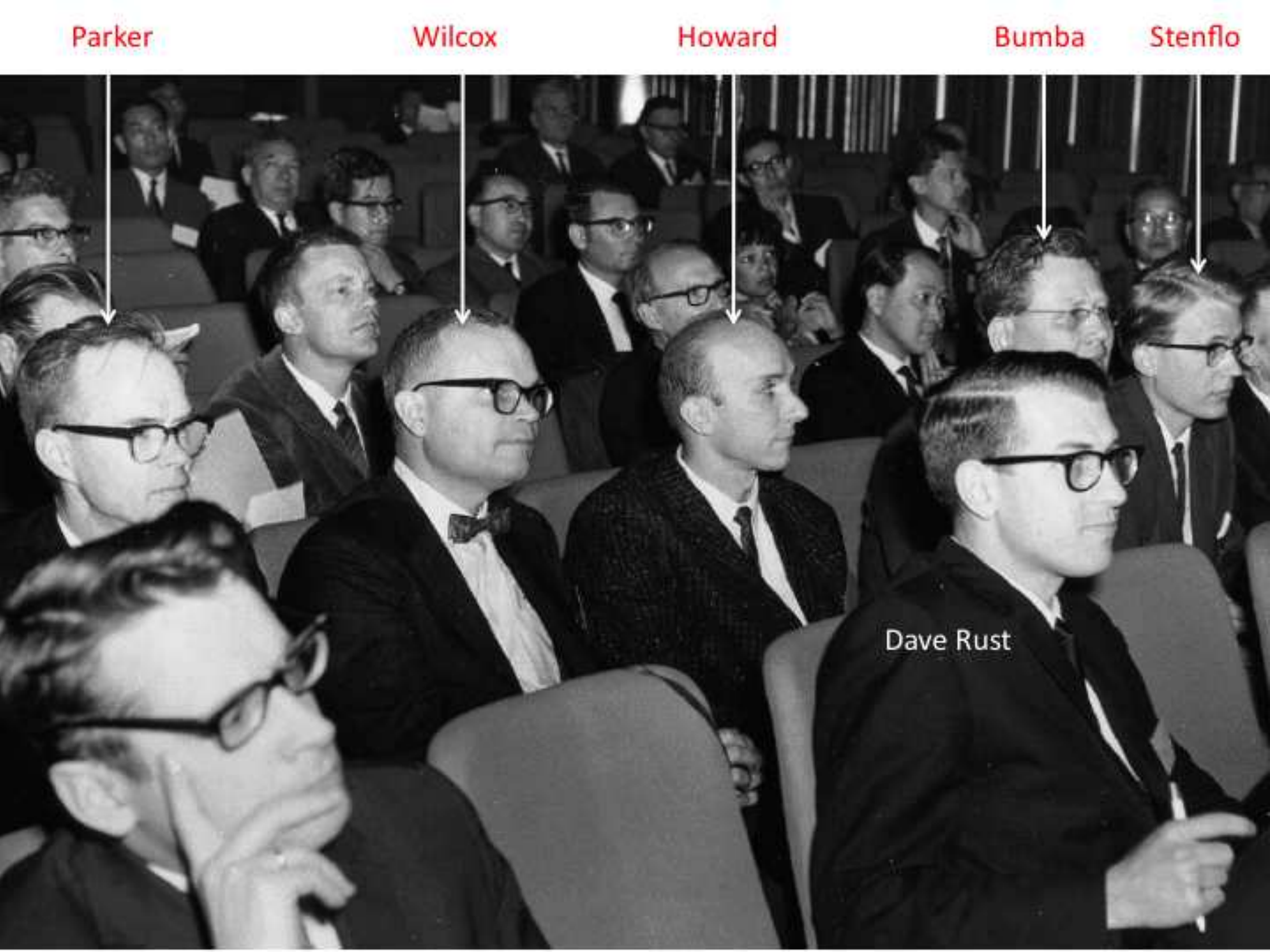}
\caption{Eugene Parker (left), here at the 1968 COSPAR meeting in
  Tokyo, made seminal contributions in theoretical solar physics over
  a wide range of topics. }
\label{fig:cospar} \end{figure*}

While Walter Elsasser from the University of Utah had overcome this
obstacle and established the mathematical framework of dynamo
theory, and in particular applied it to explain terrestrial magnetism
\citep{stenflo-elsasser1946,stenflo-elsasser1955}, Eugene 
Parker of Chicago extended the ideas to the solar
case and highlighted how Coriolis forces in a rotating and
turbulent medium make the turbulence cyclonic and thereby can
regenerate the poloidal field from the toroidal one
\citep{stenflo-parker55}. Babcock made use of these ideas 
when he built his phenomenological model of the solar cycle. Parker
was the leading theoretical solar physicist for several decades, with
fundamental contributions over a broad range of topics, including the
origin of the solar wind, the cause of intermittency in the photospheric
magnetic flux, and the mechanism for coronal heating. Figure
\ref{fig:cospar} shows him at the COSPAR meeting in Tokyo in 1968. 

%%%%%%%%%%%%%%%%%%%%%%%%%%%%%
\subsection{Mean-field MHD and the $\alpha$--$\omega$
  dynamo}\label{sec:meanfield}
\begin{figure*}\sidecaption
\includegraphics[width=0.65\textwidth]{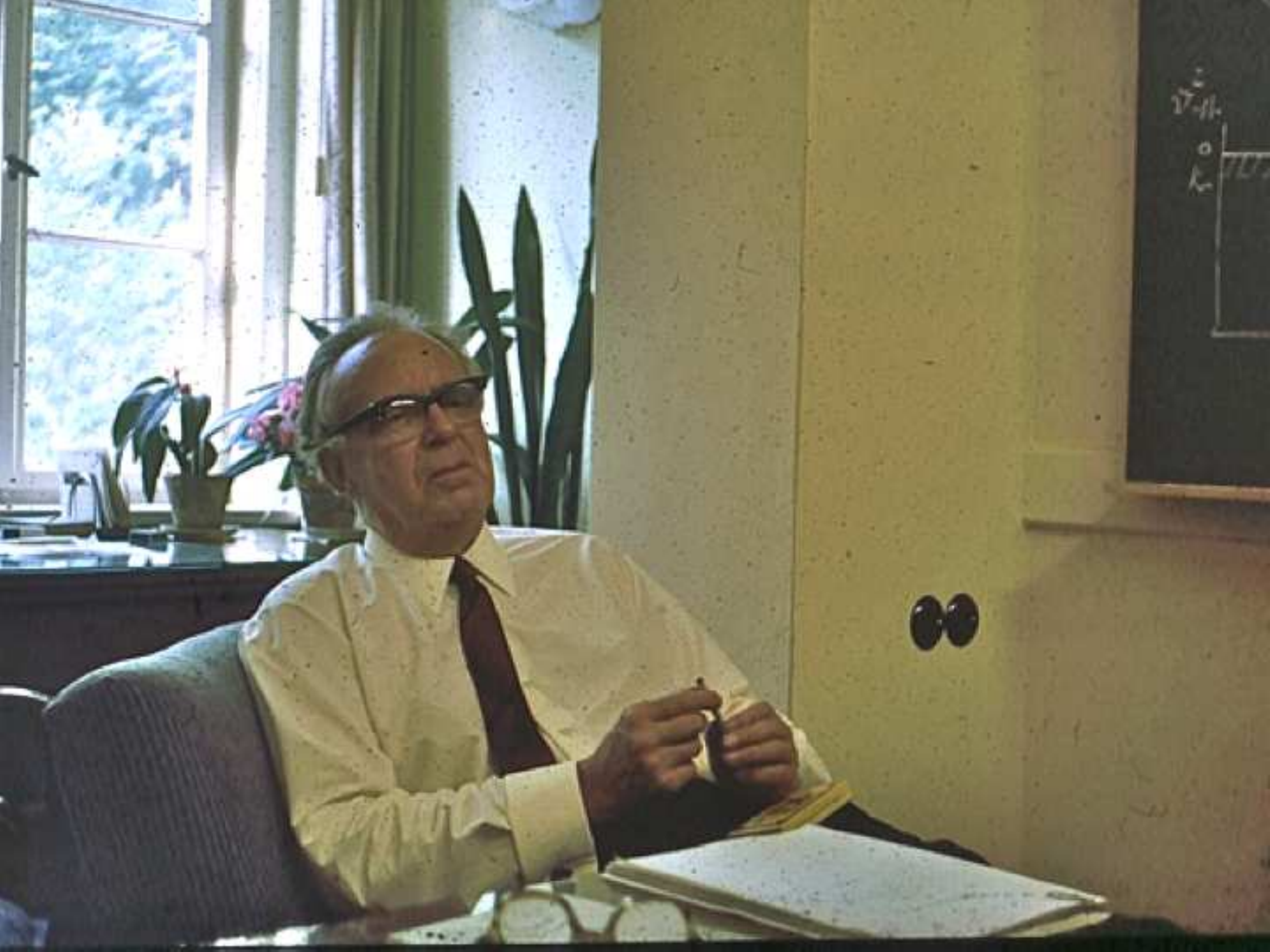}
\caption{Max Steenbeck in his office in Jena in 1969. He was the
  ``father'' of mean-field electrodynamics and $\alpha -\omega$ dynamo theory.}
\label{fig:steenbeck} \end{figure*}

Alfv\'en had heard that Max Steenbeck in Jena, DDR, had developed a
new and comprehensive formulation of the dynamo problem, and therefore
sent me to Jena in 1969 to find out how the theory worked. Max
Steenbeck was a leading person in the East German establishment, head
of the country's research council with the rank like that 
of a government minister. However, he was foremost an 
innovative scientist, who had built up a leading group in plasma
physics in Jena. Soon after my arrival he took me into his office,
called in his young collaborators Fritz Krause and Karl-Heinz
R\"adler, and described to me on the blackboard the ideas behind the
$\alpha -\omega$ dynamo. Figure \ref{fig:steenbeck} shows Steenbeck at
his desk on that occasion in 1969.  

The key to a comprehensive mathematical formulation of the dynamo
problem was the development of mean-field electrodynamics
\citep{stenflo-steenkrause69}.  Assuming
that the spatial and temporal scales of the turbulence are small in
comparison with the global scales to allow ensemble avergaging, one
could express the statistical effect of cyclonic turbulence as an
electromotive force along the field lines with $\alpha$ as the
proportionality constant. A second-order term in the ensemble average
of the turbulence represented turbulent diffusion with coefficient
$\beta$. The poloidal field lines get wound up by
the shear (radial and latitudinal) in the Sun's rotational angular velocity
$\omega$. The evolution of the global solar magnetic field can then  be
formulated in terms of two coupled equations describing the interplay
between the $\alpha$, $\beta$, and $\omega$ effects, one equation for
the poloidal component, one for the toroidal. Decomposition of these
components in their spherical harmonics leads to a coupled system of
equations for the eigenvalue problem, which allows very explicit and 
consistent modeling of the solar cycle. This type of modeling is very
general and can be applied to account for the magnetic fields in various types of
other stars, planets, and galaxies. 

I returned to Jena in 1970 at the time when the institute was being
moved from Jena to Potsdam, and I participated in the move. The Potsdam
institute established itself as a leading center for dynamo research
during the following decades. 

%%%%%%%%%%%%%%%%%%%%%%%%%%%%%
\subsection{Synoptic programs}\label{sec:synoptic}
In the caption to Fig.~\ref{fig:cowling} we mentioned how Bob Howard took over the
pioneering work of Horace Babcock and developed the first synoptic
program for full-disk line-of-sight component magnetograms recorded
daily at the Mount Wilson Observatory. Later (in the early 1970s)
similar synoptic programs 
were started at Stanford by John Wilcox and at Kitt Peak by Bill
Livingston and Jack Harvey. The Stanford program continues to this day
with well defined but low spatial resolution. The Kitt Peak program has
had the highest spatial resolution of the ground-based programs and
was upgraded in the late 1990s by 
Jack Harvey and Christoph Keller as part of the SOLIS synoptic
program. 

The most advanced synoptic full-disk magnetogram projects have been
the space programs {\it SOHO}/MDI (1995-2011) and its superior 
successor, {\it SDO}/HMI. Both have been managed by the Stanford group. 

\begin{figure*}\sidecaption
\includegraphics[width=0.5\textwidth]{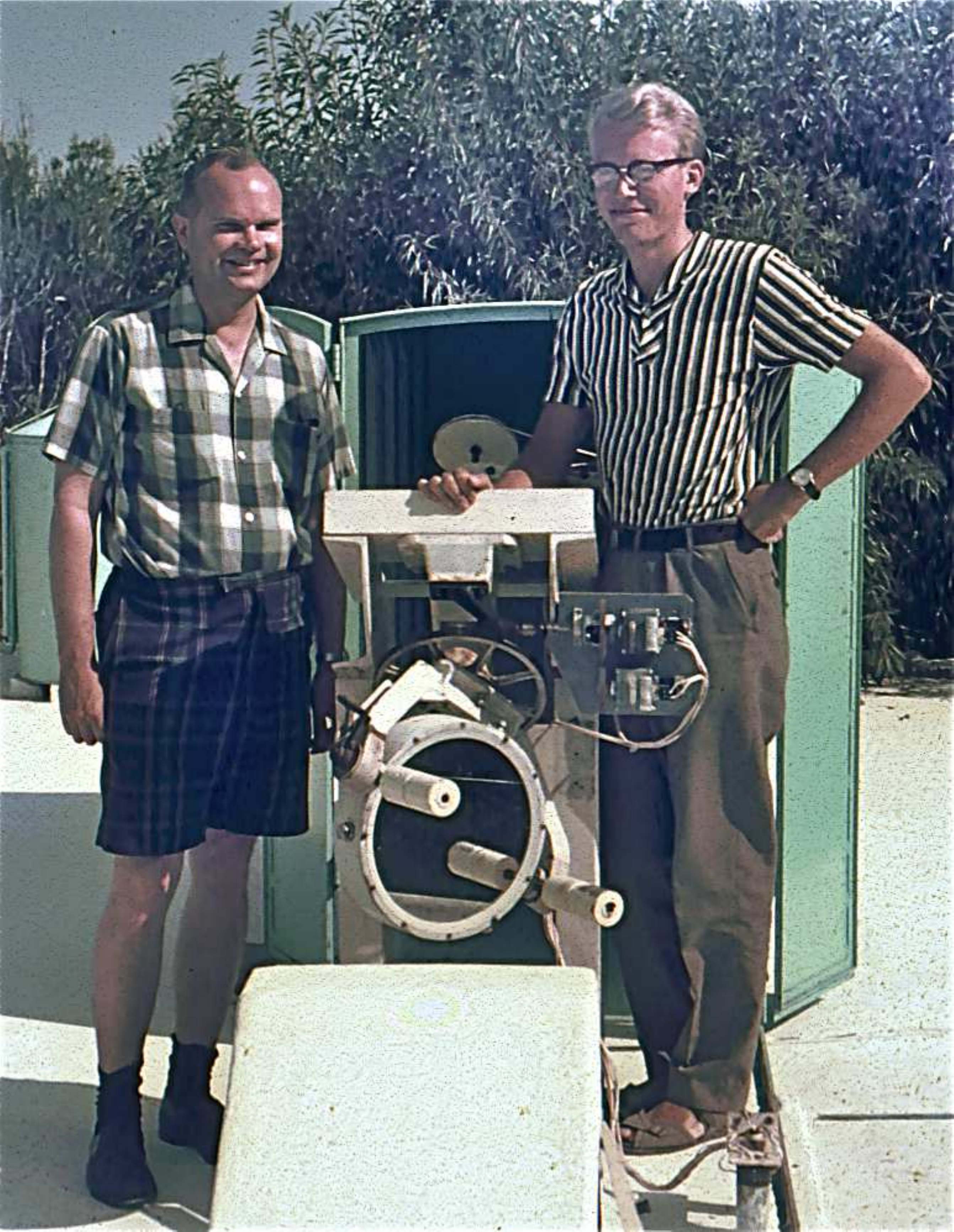}
\caption{John Wilcox (left) with Jan Stenflo in 1967 at the Swedish
  solar station on Capri, Italy. Wilcox was a leading solar wind
  scientist and founded the synoptic full-disk magnetogram program at
  the Stanford University.}
\label{fig:wilcox} \end{figure*}

John Wilcox had close contacts with Hannes Alfv\'en and was in the
audience in Stockholm when I in 1966 gave a colloquium at Alfv\'ens
institute on my observations of the Sun's magnetic fine structure at
the Crimean Astrophysical Observatory. His main scientific focus at
the time was on the physical properties of the solar wind and its
sources on the Sun. He visited me at the Swedish solar station on
Capri in 1967 (see Fig.~\ref{fig:wilcox}). When I for the first
time set foot on US soil (in 1968 after the defense of my
Doctoral dissertation in Lund) Wilcox had arranged that I spent the
first four nights at the Faculty Club of the Berkeley
campus (later Wilcox moved from Berkeley to
Stanford). I had made a world-around trip, so I entered the US from
the west. It was Berkeley June 1968 at the height of the Vietnam war,
and it was my first impression of what American society was like\,! 

After Wilcox' tragic accidental death in 1984 by drowning while swimming in
Hawaii, the synoptic Stanford Solar Observatory was renamed Wilcox
Solar Observatory.

%%%%%%%%%%%%%%%%%%%%%%%%%%%%%
\section{The quest for angular resolution}\label{sec:resolution}
When the Sun is observed with improved angular resolution we discover
previously unknown structures which deepen our
physical understanding, like the granular convection pattern, the
chromospheric filamentary structures, network and internetwork bright
points, etc. Photospheric magnetic flux is highly intermittent with
basic kG flux elements having typical sizes of order 100\,km or
below. Most of the size distribution is beyond reach of current
telescopes, although flux tubes in the larger-scale tail of the
distribution have been resolved. The magnetic-field pattern has a
fractal-like structure, with a structuring that continues far beyond the
resolution limit. 

It has therefore always been a high priority for solar physicists to
push for ever higher angular resolution, since it is widely believed
that the fundamental physical mechanisms to a large extent are
connected to processes at the smallest scales.

%%%%%%%%%%%%%%%%%%%%%%%%%%%%%
\subsection{JOSO and LEST}\label{sec:JOSO}
Half a century ago the leading solar physics facilities were in the US (Mount
Wilson in California, Kitt Peak with the McMath telescope in Arizona,
Sacramento Peak in New Mexico) and the USSR (Crimean Astrophysical
Observatory). The European solar physics community felt a strong need
to overcome the political fragmentation after two devastating world
wars and create a joint European solar facility that
would again give Europe a leading role in science, as had been done
by CERN for particle physics and by ESO for night-time astronomy. The
driving force behind this pan-European solar physics movement was
Karl-Otto Kiepenheuer, Director of the Fraunhofer Institute (after his
death renamed Kiepenheuer Institute for Solar Physics) in
Freiburg, Germany. In the late 1960s Kiepenheuer set up an informal
European organization with the name JOSO, Joint Organization for Solar
Observations \citep{stenflo-kiepenheuer73}, which had the goal of
constructing a major European 
solar telescope, LEST (Large European Solar Telescope). This goal was
to be achieved in two steps: (1) Search by systematic site
testing campaigns for the optimum location of LEST. (2) Establishment of a new
organization with legal structure and adequate financial means to
design and construct LEST at the chosen site. 

\begin{figure*}
\centering
\includegraphics[width=0.95\textwidth]{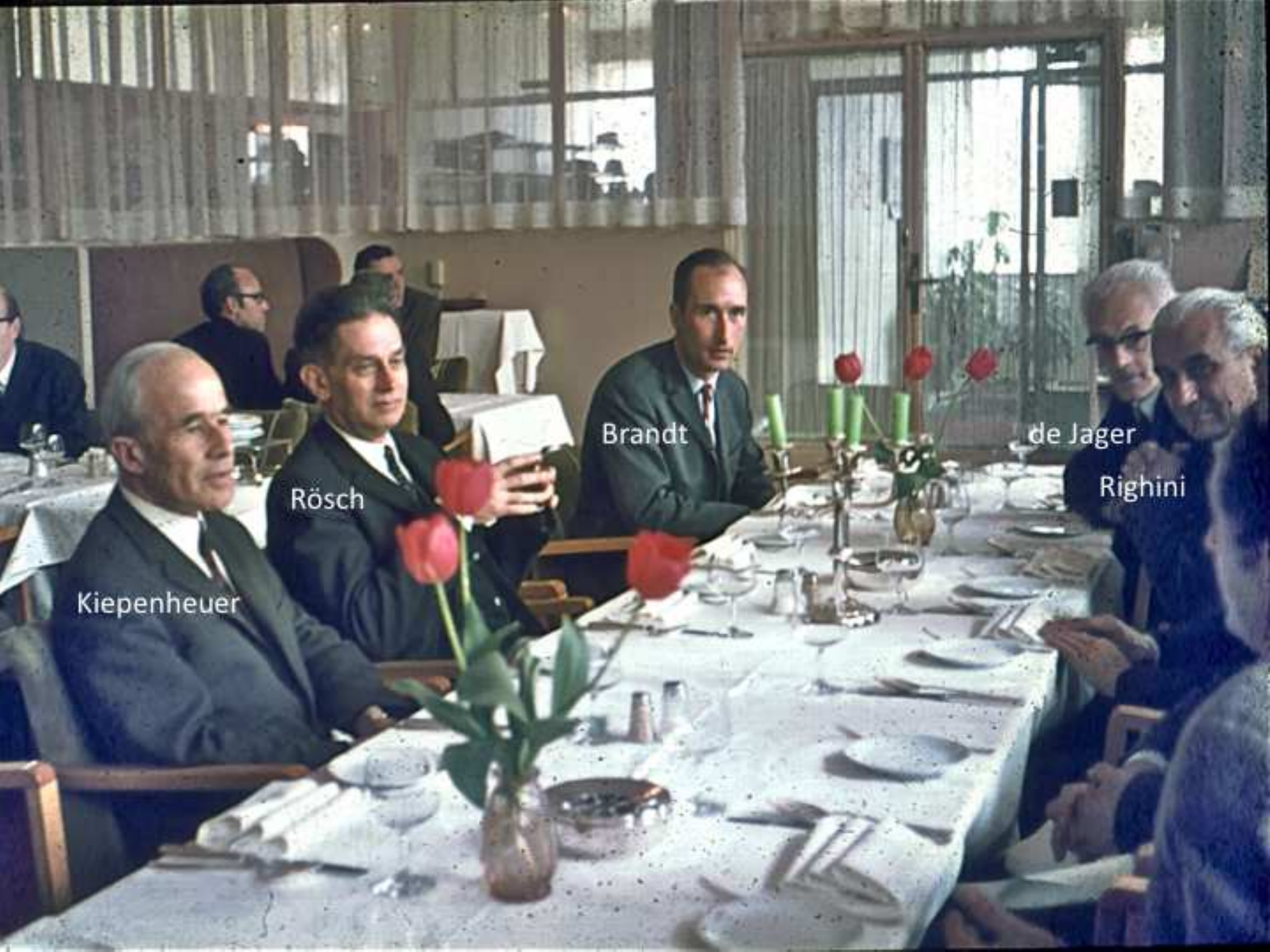}
\caption{JOSO meeting 1969 in Freiburg, Germany. From left to right:
  Karl-Otto Kiepenheuer (Freiburg), Jean R\"osch (Pic du Midi), Peter
  Brandt (Freiburg), Kees de Jager (Utrecht), Guglielmo Righini (Arcetri).}
\label{fig:kiepen} \end{figure*}

Figure \ref{fig:kiepen} shows Kiepenheuer at one of the first meetings
of JOSO, in Freiburg 1969. Beside him is Jean R\"osch, Director of the
Pic du Midi Observatory, representing France. Holland was represented
by Kees de Jager, who founded the journal Solar Physics and later
served as president of COSPAR, Italy by Guglielmo Righini,
Director of the Arcetri Observatory in Florence, while I was representing Sweden. 

After having done site testing of a number of coastal sites in
Portugal, Sicily, and Greece it was concluded that high-altitude
volcanic island sites would be better, and that therefore the site
testing should concentrate on Tenerife (Iza\~na) and La Palma (Roque de
los Muchachos) in the Canary Islands. Towards the end of the 1970s
these site testing campaigns came to the conclusion that both Iza\~na 
and Roque de los Muchachos were indeed outstanding sites of comparable
quality. There was no clear recommendation which of the two was
better. Both are at the same elevation, just above the well-defined
inversion layer where the clouds form. At these sites the wind
direction is most of the time such that the line of sight from the
telescope to the Sun is in a laminar air flow if the telescope sits on
top of a tower high enough to avoid ground turbulence. This leads to
excellent seeing, while the high altitude causes scattered light from
the sky to be low. 

Rather than waiting for a LEST to be realized, various countries
quickly decided to set up their national facilities on these
sites. Germany chose Iza\~na, where the Kiepenheuer Institute in
Freiburg built the VTT (Vacuum Tower Telescope) while the University
Observatory of G\"ottingen set up the Gregory-Coud\'e\ telescope that
they had transferred from their solar station in Locarno,
Switzerland. Similarly the French built their THEMIS 90\,cm aperture
polarization-free telescope at Iza\~na. The Swedes (through their
observatory director Arne 
Wyller, who had succeeded Yngve \"Ohman, founder of the Swedish
Capri observatory) transferred their solar station in Anacapri to Roque de los
Muchachos, where also the Dutch established themselves with the DOT
(Dutch Open Telescope), designed by Rob Hammerschlag. Soon after, the
telescope at the Swedish La Palma observatory was completely rebuilt
with an innovative design by G\"oran Scharmer, who later took over as
observatory director. Under Scharmer
the La Palma facility became the unrivaled world leader in solar
imaging with the highest angular resolution. No other solar facility
in the world has yet consistently achieved such resolution. 

In parallel with these national activities the efforts towards the
realization of the LEST dream made rapid progress. A new organization,
the LEST Foundation, was established in 1983 with legal seat at the
Royal Swedish Academy of Sciences. Its 
task was to execute the intentions of JOSO, to obtain the financial
means, develop the detailed design, and to construct and operate LEST
at the chosen site. The scientific aim was focused on recording
magnetic fields with the highest possible spatial
resolution \citep{stenflo-lest85}. Throughout its existence the LEST Foundation was 
led by a ``triumvirate'', with me as President, Oddbj\o rn Engvold of Oslo
as Project Director, and Kai-Inge Hillerud of the Royal Swedish
Academy in Stockholm as Executive Secretary. The innovative
design was led by Engvold with crucial input from Richard Dunn, US
master optical designer, who made the 
Sacramento Peak Observatory (New Mexico) become the leading
high-resolution facility in the US, and from Torben Andersen, who was
a Danish telescope designer at ESO. Since other non-European countries like
the US and China joined the LEST Foundation as members, the acronym
was redefined as Large Earth-based Solar Telescope, to emphasize that
it was not just a European but a truly international undertaking. 

Around the
end of the 1980s the detailed design for the entire facility had been
completed, the definite site had been selected (on Roque de los
Muchachos near the Swedish La Palma observatory), we were ready to start the construction
phase, and it appeared that the needed funding support was
forthcoming \citep{stenflo-engvold91}. Dramatic external events
changed the situation. When the 
Soviet Union collapsed, Germany turned all its priorities and
financial resources toward the enormous task (and historical
opportunity) of German reunification, with the consequence that the
expected substantial German contributions to the construction phase of
LEST would not be available in the foreseeable future. The LEST
project was put on hold for another decade (although during this time
many design studies for various focal-plane instrumentations were
carried out), until it was finally decided to dissolve the LEST
Foundation as of June 30, 2002. 

As the LEST project lost momentum in the 1990s, the idea of a
next-general type international solar facility was being developed in
the US in the form of ATST (Advanced Technology Solar Telescope), with
an innovative 4\,m aperture telescope design by Jacques Beckers. This
ambitious project, renamed DKIST (Daniel K. Inouye Solar Telescope) is
now in construction at the Haleakala site on Maui and is expected to
be in operation by 2020. 

Before the LEST Foundation was dissolved, Torben Andersen and Oddbj\o rn
Engvold designed a scaled down version of LEST, which has since been
implemented in the form of 
the Chinese 1-m solar telescope at the Fuxian lake near Kunming and
the 1.5-m German GREGOR telescope on Tenerife (which has replaced the
45\,cm G\"ottingen Gregory-Coud\'e telescope). The original LEST idea
of a joint European solar facility has been resurrected within Europe
in the form of EST (European Solar Telescope), aiming for a telescope in the
DKIST class, and it is progressing well with the help of EU funding.

%%%%%%%%%%%%%%%%%%%%%%%%%%%%%
\subsection{The McMath-Pierce facility as a center of
  innovation}\label{sec:mcmath} 
In the late 1960s and the following decades the leading US
facility for solar observations with high spatial resolution was the
tower telescope at the Sacramento Peak Observatory in New Mexico,
which had been designed by Richard Dunn and later given the name DST (Dunn
Solar Telescope). However, the main innovations and advances in the
exploration of solar magnetic fields took place at the McMath
telescope at Kitt Peak because of the many contributions of in
particular William Livingston,  Jack and Karen Harvey, Jim Brault, and Neil
Sheeley. This giant telescope had been designed by Robert McMath but
was renamed the McMath-Pierce facility after the death of the head of
the Kitt Peak solar department, Keith Pierce. 

\begin{figure*}
\centering
\includegraphics[width=0.85\textwidth]{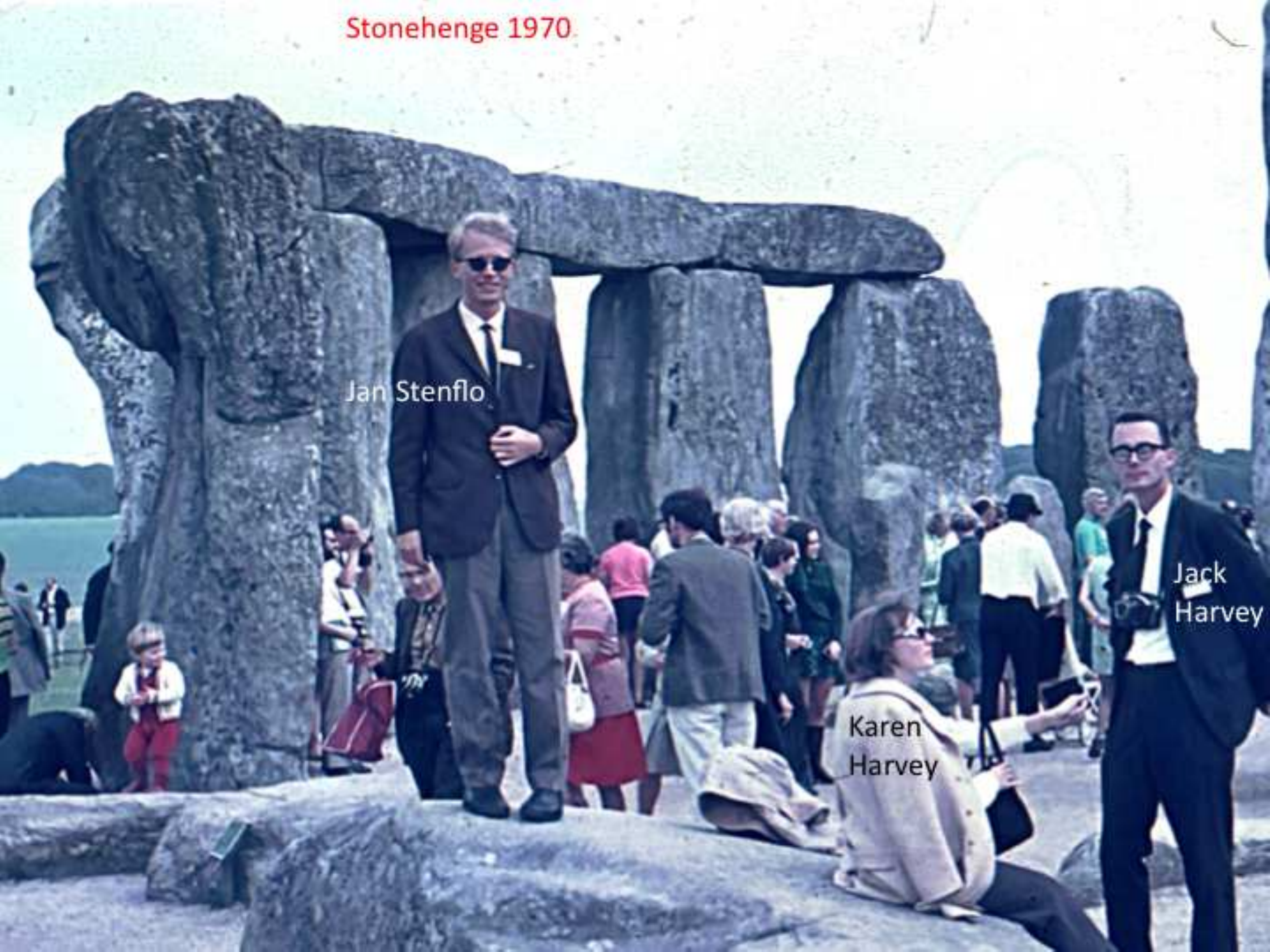}
\caption{Jack and Karen Harvey with Jan Stenflo at Stonehenge,
  1970. Jack did ground-breaking work in the 1970s on the small-scale
  structure of solar magnetic fields, and Karen codiscovered the
  ephemeral active regions.}
\label{fig:stonehenge} \end{figure*}

Harvey and Livingston implemented innovative techniques for
magnetic-field mapping, which much advanced our insight and
understanding of the small-scale nature of the fields
\citep[cf.][]{stenflo-livingstonetal76,stenflo-harvey77}. Around 1974 
they initiated the synoptic program of daily Kitt Peak full-disk
magnetograms and synoptic maps that resolved the field in much greater
detail than anywhere else. Jack's wife Karen discovered together with
Sara Martin of the San Fernando Observatory the ephemeral active
regions \citep{stenflo-harveymartin73}. Jim Brault developed a Fourier
transform spectrometer (FTS) 
for the visible part of the spectrum
\citep{stenflo-brault78,stenflo-brault85}, which with the help of Jack 
Harvey could be converted into an FTS polarimeter. 

My own main observational work has been done with the McMath-Pierce
facility. In September 1971 I applied the magnetic line-ratio
technique (with the 5250-5247\,\AA\ line pair) to discover that more
than 90\,\%\ of the magnetic flux that had been recorded in solar
magnetograms (with a resolution of a few arcsec) on the quiet Sun
actually is in kG form but appears weak since it is unresolved with a
small filling factor \citep{stenflo-s73}. In 1978-79 I used the FTS polarimeter to make
atlases of the Stokes $V$ spectra in plages and the network and
surveys of the linearly polarized spectrum near the solar limb. The
Stokes $V$ spectra \citep{stenflo-setal84} led to detailed physical validation of the
line-ratio technique, revealed the nature of the Stokes $V$
asymmetries, and could be used by Sami Solanki, my first PhD student
at ETH Zurich, to apply multi-line techniques
for the empirical modeling of solar magnetic flux tubes. The linear
polarization survey led to the discovery of the Second Solar Spectrum
(cf. Sect.~\ref{sec:ss2}), 
which is formed exclusively by coherent scattering processes and 
is the playground for the Hanle effect. When the ZIMPOL technology had
been developed at ETH Zurich it was used from 1994 onwards at the
McMath-Pierce facility for systematic explorations of the Second Solar
Spectrum with a polarimetric precision of $10^{-5}$. The main
Hanle-effect constraints on the properties of the hidden turbulent
magnetic fields have come from ZIMPOL observations at Kitt
Peak. 

The main reason why the McMath-Pierce facility has served as the
premier place for innovative, experimental solar science is, besides
its high light-gathering capacity (1.5\,m entrance aperture) the
large, open experimental environment in the spatious observing room,
with free access to the large solar image and unrestricted access to
the beam in front of the entrance slit of the spectrograph, which
makes it simple to insert novel experimental equipment. Other large astronomical
telescopes generally do not allow experimental freedom of this
kind. It will be greatly missed with the closure of the McMath-Pierce
facility.

%%%%%%%%%%%%%%%%%%%%%%%%%%%%%
\subsection{FTS polarimetry}\label{sec:fts}
The FTS polarimeter allowed unprecedented insight into the physical
nature of the spatially unresolved magnetic fields. It allowed
the simultaneous recording of the spectra of Stokes $I$ (the ordinary
intensity) and one of $Q$, $U$, or $V$ with full spectral
resolution (absence of significant instrumental spectral smearing), no
spectral stray light, and with simultaneous wavelength coverage over
the full spectral range admitted by the broad-band prefilter used
(typically 1000\,\AA). This was far beyond what one could dream of
achieving with any spectrograph system. The trade-off was low angular
resolution (5\,arcsec at the time) and temporal resolution (tens of
minutes). Since the vast majority of solar physicists had their focus
on high spatial resolution, few took interest in the use of the FTS
polarimeter, I became almost its only user.

\begin{figure*}\sidecaption
\includegraphics[width=0.6\textwidth]{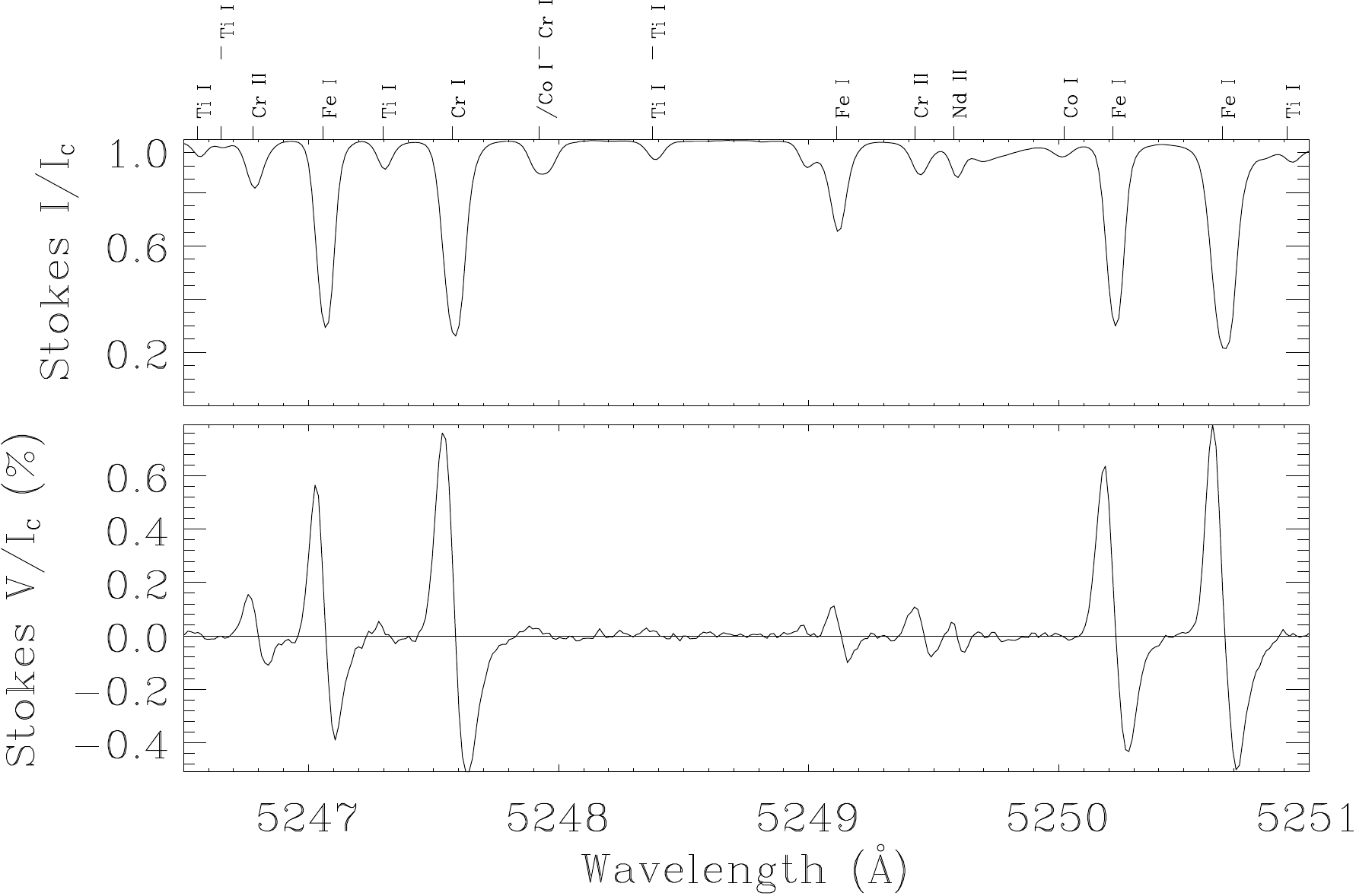}
\caption{Illustration of a small spectral window from the Stokes $I$
  and $V$ atlas of a weak plage, recorded with the FTS polarimeter in
  1979 \citep{stenflo-setal84,stenflo-s13aarv}.}
\label{fig:fts} \end{figure*}

As an example Fig.~\ref{fig:fts} shows a 4.5\,\AA\ wide portion of the
FTS spectral atlas of a weak plage recorded in 1979. In particular it
illustrates the exact behavior of the Stokes $V$ profiles of the lines
Fe\,{\sc i} 5247.06 and 5250.22\,\AA, which belong the magnetic line-ratio
pair that was used with the McMath Babcock-type magnetograph system to
reveal that most of the photospheric magnetic flux that had been recorded in
magnetograms has its origin
in strong, 1-2\,kG bundled fields with small filling factors. Analysis
of the observed Stokes $V$ 
line shapes allows a detailed validation of this interpretation. The
FTS Stokes $V$ atlases are available both as pdf and data files at 
http://www.irsol.ch/data\_archive/\#ftsv .  

The Stokes $V$ spectral atlases and their center-to-limb variations
represented a gold mine for the construction of empirical models of
the spatially unresolved kG flux tubes, which was exploited by Sami
Solanki in his PhD thesis project at ETH Zurich. Sami went on to become
director at the Max-Panck-Institute for Solar System Research in
G\"ottingen, where he led the successful {\it Sunrise} project 
to observe solar magnetic fields with unprecedented spatial resolution
from a stratospheric balloon \citep{stenflo-sunrise2010}. 

The FTS polarimeter represented a remarkable advance in solar
instrumentation, but it has now been decommissioned, and no similar
instrument exists anywhere else. There are not even plans within any
of the main solar telescope projects to build another instrument of
this type. Like the 1970s represented a heroic era when we could fly
people to the moon but are not able to do it any more (or have lost
interest in it), FTS polarimetry represents a lost art, without
sufficient motivation anywhere to resurrect it. 

It is true that the FTS polarimeter in the 1970s had low spatial and
temporal resolution and used 1-pixel detectors, but it was in
principle possible to
improve the resolutions and to use multi-pixel detector
arrays, although it is technically difficult. However, these
improvements were never done, because they were not requested by the
broader solar community.

%%%%%%%%%%%%%%%%%%%%%%%%%%%%%
\subsection{The endless debate on the existence of kG
  fields}\label{sec:kg}
Application of the magnetic 5250/5247 line ratio method through my 1971
Kitt Peak observations allowed me to conclude that most of the
measured quiet-sun magnetic flux is in kG
form \citep{stenflo-s73}. For me the evidence was so clear that I
considered the case as 
closed. Since however this conclusion was not reached by 
resolving the kG fields, but indirectly, through the identification of a telltale spectral
signature that could not have any other interpretation, it took
many years before my result found broad acceptance within the
community, even after the interpretation got validated
with the independent FTS polarimeter data. Over the years various
people presented their own 
conclusions about the existence of quiet-sun kG fields as if it 
were a new discovery.  

While as time went by ever fewer people questioned the existence of kG flux
on the quiet Sun, two leading scientists within the solar community never
accepted it: Andrei Severny, director of the largest astrophysical observatory in
the Soviet Union, and Harold Zirin of Caltech, who founded
the Big Bear Solar Observatory (which presently has the largest-aperture solar
telescope in the world). After Severny passed away
in 1987, Hal Zirin remained the stronghold of resistance. 

Hal seemed to enjoy this controversy. In 1992 he organized an IAU
Colloquium in Beijing and invited me to have a debate with him there
on the reality of the kG
fields \citep[cf.][]{stenflo-s93debate,stenflo-z93debate}. The debate
between the two of us was 
set up to represent a highlight of the conference. The lively event 
did not change any opinions, but I remember the
atmosphere as friendly and entertaining. Hal and I respected each
other in spite of our opposite views. 

Long after, it seems that some remember this event as a debate
about whether the quiet-sun fields are weak or strong. This is not at
all what it was about. There has always been agreement 
that most of the photospheric volume is filled with weak fields. The
intermittent kG flux picture that resulted from application of the
line-ratio technique implied that only about 1\,\%\ of the
photospheric volume is occupied by kG fields, while the remaining
99\,\%\ has much weaker fields. The discovery of the kG fields 
was followed by a focus on finding the properties of the remaining
99\,\%\ weak fields. These efforts led to the discovery that one can
use the symmetry
properties of the Hanle effect to constrain the field strengths of these
weak fields \citep{stenflo-s82}. 

So the Beijing debate in 1992 was not about whether the fields are
strong or weak, it was whether there exist kG-type magnetic fields on
the quiet Sun, which carry much of the magnetic flux 
that we see in solar magnetograms. Zirin always denied the existence
of such fields, and he never gave up this position.

%%%%%%%%%%%%%%%%%%%%%%%%%%%%%
\subsection{Ensemble averages and angular
  distributions}\label{sec:ensemble}
While the initial observations of quiet-sun magnetic fields by Babcock
was with arcmin angular resolution, the observing techniques advanced
rapidly in subsequent years, so that by the end of the 1960s an
angular resolution of better than a few arcsec could be reached. With
each new resolution breakthrough there was a tendency to proclaim that
the magnetic structures had finally been resolved. We now know that
such proclamations were wildly premature, since the magnetic
structuring continues to far smaller scales (cf. Fig.~\ref{fig:enspec}
below). 

With the realization that the fields remain largely unresolved came
the need to find resolution-independent spectral signatures that could
reveal the intrinsic properties of the fields as they would appear with
infinite resolution. Since the field morphology is not resolved, such
techniques must be statistical in nature, making use of ensemble
averages of field elements to find the intrinsic statistical 
properties of the elements that make up the ensemble. The magnetic
line-ratio method is an example of such a technique. Since the field
elements are far from resolved, the angular-resolution window of the
instrument averages over an ensemble of fields with a vast range of field
strengths, both weak and strong. The {\it differential} effects
between the ensemble-averaged Stokes $V$ profiles of the
5250-5247\,\AA\ line pair allow us to conclude that most 
of the resolved Stokes $V$ signal gets its contribution
from intrinsically kG flux elements. The 5250-5247 pair is the only known
line combination for which the kG interpretation is unique, because 
thermodynamic and radiative-transfer effects are unable to produce
differential effects of the observed kind (in contrast to the case for
all other considered line combinations). 

Two other types of spectral signatures from ensemble
averages of Stokes parameters have led to profound insights into the
hidden small-scale nature of solar magnetic fields: (1) Depolarization
signature in the linear scattering 
polarization, caused by the Hanle effect of an ensemble of
microturbulent (optically thin) magnetic fields with random
orientations of their field vectors \citep{stenflo-s82}. These fields
represent most of the
weaker (non-kG) fields that occupy approximately 99\,\%\ of the
photospheric volume. (2) Sign pattern of the ensemble-average of Stokes $Q$
from the transverse Zeeman effect as observed away from disk center
(which is needed to break the symmetry to give non-zero
signatures). This sign pattern is an unbiased and
resolution-independent signature of the vertical or horizontal
preference (with respect to the isotropic case) of the 
angular distribution of the field vectors of the ensemble
\citep[][cf. the discussion at end of Section \ref{sec:vector}]{stenflo-s87}.

%%%%%%%%%%%%%%%%%%%%%%%%%%%%%
\section{Hanle effect and the Second Solar Spectrum}\label{sec:hanless2}
Since Hale the Zeeman effect has been the main tool for the
diagnostics of cosmic magnetic fields. In recent decades, however, the
Hanle effect has emerged as a  
complementary tool that is sensitive to fields in parameter domains
that are nearly inaccessible to the Zeeman effect. The main
reason why it has not yet been so widely used but
rather been an area for a smaller community of specialists is that the
underlying physics is complicated and the applications are not so
straightforward. The observations require higher polarimetric
sensitivity, and the interpretations generally need vector radiative
transfer with scattering and partial frequency redistribution in
magnetized media, a subject area that is still not fully
developed. The interpretations are also sensitive to the assumptions
concerning the 3-D geometry of the atmosphere. Due to such
complications much of the astrophysical exploitation of 
the Hanle effect still lies in the future. 

\begin{figure*}\sidecaption
\includegraphics[width=0.6\textwidth]{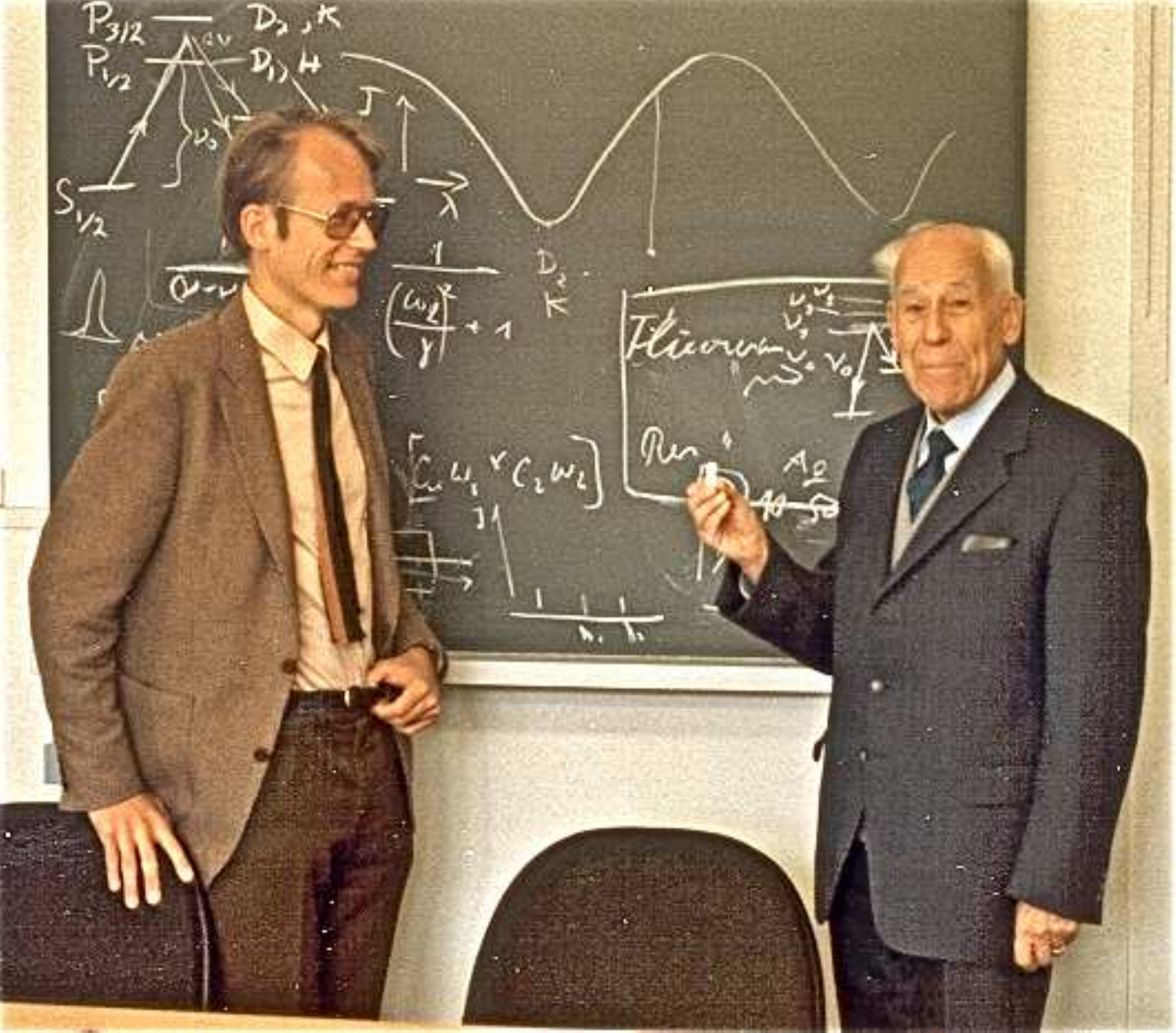}
\caption{Wilhelm Hanle visiting me in my ETH office in 1983 on the
  occasion of the 60th anniversary of his effect.}
\label{fig:hanle} \end{figure*}

The Hanle effect was discovered by Wilhelm Hanle
(cf. Fig.~\ref{fig:hanle}) in 1923 during his PhD work in G\"ottingen
\citep{stenflo-hanle24}. It played an important role in the early development of
quantum mechanics because it represented an explicit expression of
the concept of linear superposition of quantum states 
and the decoherence that is caused by an external magnetic field. It
soon found applications over a wide range of fields in physics before
it was realized that it could also have important applications in
astrophysics \citep[cf.][]{stenflo-mor91}. Hanle himself was
enthusiastic when he learnt about the astrophysical applications.

%%%%%%%%%%%%%%%%%%%%%%%%%%%%%
\subsection{Coherent scattering and Hanle effect on the Sun}\label{sec:cohere}
The blue sky is polarized due to Rayleigh scattering at molecules in
the Earth's atmosphere. Similarly the Sun's spectrum gets linearly
polarized because coherent scattering processes in the continuum and
the various spectral lines contribute to the formation of the
spectrum. Magnetic fields are not the source of the scattering
polarization, but they modify it and thereby generate observable
spectral signatures. It is this magnetically-induced modification that
is referred to when we speak of the Hanle effect. 

It was realized already in the 1920s by \citet{stenflo-ohman29} that
such scattering polarization 
should exist on the Sun. Yngve \"Ohman was the
founder of the Swedish solar station on Capri and was one of my
mentors (since I started my solar work 
on Capri at age 20 in 1963). The first reliable
observational evidence came when G\"unter Br\"uckner recorded the
scattering polarization near the Sun's limb in the Ca\,{\sc i}
4227\,\AA\ line with the telescope at G\"ottingen's solar station in
Locarno, Switzerland \citep{stenflo-brueckner63}. In the mid 1980s the ownership
of the observatory was transferred to Switzerland and it was
renamed IRSOL (Istituto Ricerche Solari Locarno). Its scientific focus
is the exploration of the scattering polarization and the Hanle effect
with the powerful imaging polarimetry system ZIMPOL that was developed
at ETH Zurich (see Sect.~\ref{sec:ss2}). 

\begin{figure*}\sidecaption
\includegraphics[width=0.65\textwidth]{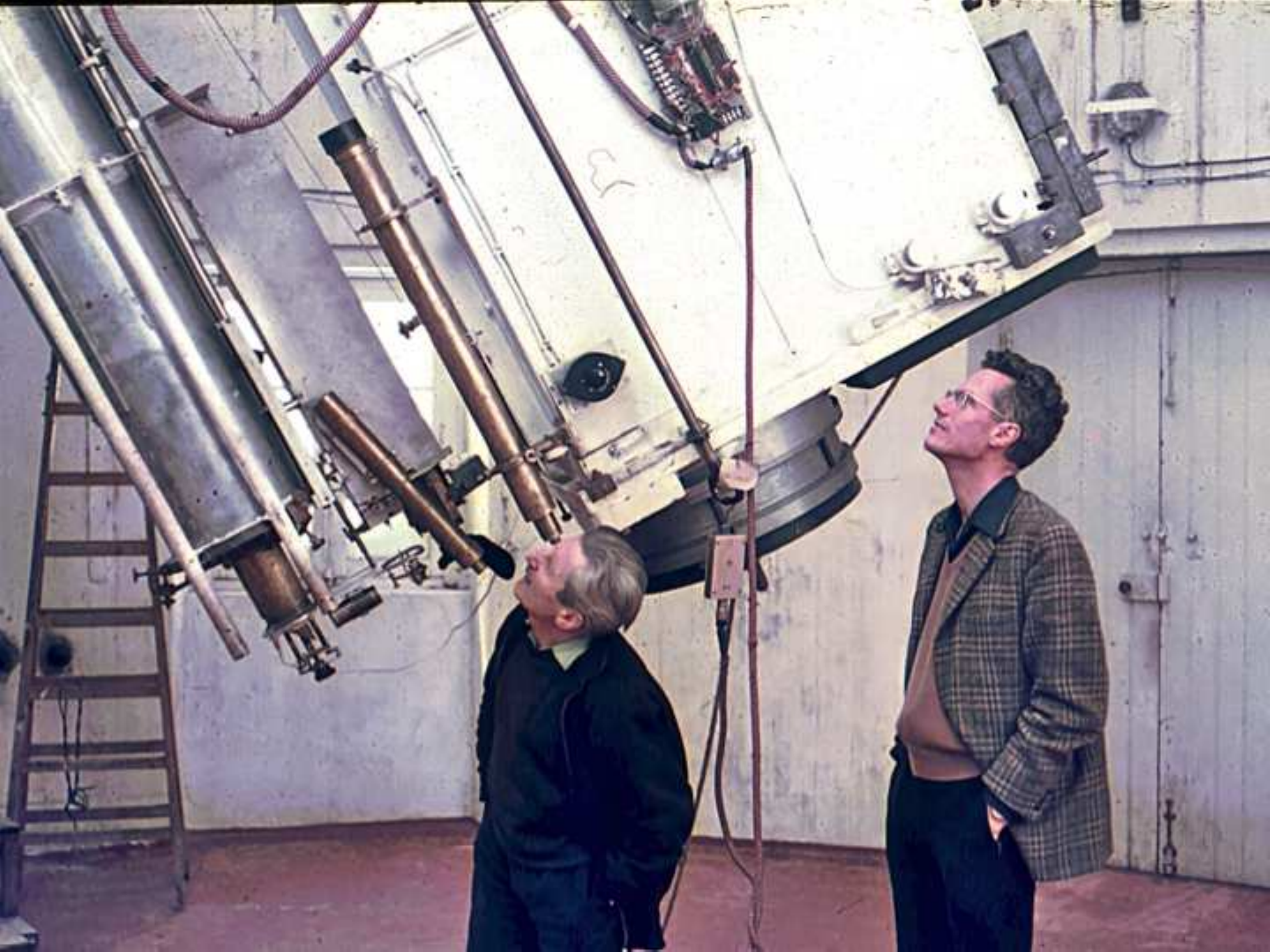}
\caption{Jean-Louis Leroy (right) at Pic du Midi, 1970.}
\label{fig:leroy} \end{figure*}

The first consistent interpretation of the non-magnetic 4227\,\AA\ observations
in terms of polarized radiative transfer was made by
\citet{stenflo-dumontetal73} in Paris. The first observation of the
Hanle effect in an 
astrophysical context was done in solar prominences by
\citet{stenflo-leroy77} at the Pic du Midi Observatory. Figure
\ref{fig:leroy} 
shows Jean-Louis Leroy when I visited him at Pic du Midi in 1970. 

The personal contacts developed during my extended periods of work in
the Soviet Union in the 1960s led to the opportunity for Sweden to
build instruments to be flown on Soviet Intercosmos satellites. Due to
the severe limitations in space, weight, 
and telemetry we had to come up with an idea of doing something
qualitatively different, not attempted with the much larger resources
available elsewhere. Our choice was to build a Swedish
spectro-polarimeter for the solar vacuum ultraviolet, to search for
signatures of polarization due to coherent scattering in H
Lyman $\alpha$ 
and in emission lines formed in the chromosphere-corona
transition region. This project represented my own entry into the
field of scattering polarization physics. 

During the first attempt to launch the Swedish experiment, in 1975, 
the Soviet rocket exploded and all the experiments were
destroyed. However, 
we had a spare model that could be successfully put in orbit the
following year on the satellite Intercosmos 16 (the satellite that
exploded was never assigned a number in the Intercosmos series ---
officially it did not exist). I
participated in the launch preparations during three weeks at 
the super-secret military launch site of Kapustin Yar near
Volgograd. It was the first time that anyone from a western country
was admitted to a Soviet launch site. 

Our UV spectro-polarimeter functioned as it was supposed to, but
contamination of the UV optics due to outgassing in orbit from other
parts of the spacecraft (because the
integration had not been performed in a cleanroom but in a dirty
garage environment) caused the optical transmission to quickly
degrade by a factor of one hundred, increasing the polarimetric noise
by an order of magnitude. It was therefore only possible to set an upper limit of
1\,\%\ for the H Lyman $\alpha$ scattering polarization near the limb 
as observed with one arcmin spatial resolution \citep{stenflo-ik80}. Although this meager
scientific output was a disappointment, the experiment demonstrated
the feasibility of doing spectro-polarimetry in the vacuum
ultraviolet. 

There is considerable potential for using the scattering
polarization and its magnetic modification via the Hanle effect in
this part of the spectrum as a tool to explore the 
elusive magnetic fields in the chromosphere-corona transition region,
where the physical processes responsible for coronal heating take
place. Still, four decades later, this potential has not yet been
exploited, but finally there is hope for some progress: the {\it
  CLASP} rocket experiment for recording the Hanle effect in H Lyman
$\alpha$ is ready to be launched \citep[cf.][]{stenflo-claspspw6}.

%%%%%%%%%%%%%%%%%%%%%%%%%%%%%
\subsection{Second Solar Spectrum}\label{sec:ss2}
The UV space experiment led to
my interest in a systematic exploration of the physics of scattering
polarization and its signatures in the Sun's spectrum. In 1978 I used
the grating spectrograph and the FTS at the Kitt Peak 
McMath telescope to make a spectral survey of the linear polarization inside
the solar limb, from 3165 to 9950\,\AA\ 
(Stenflo et al. 1983a,b). 
%\citep{stenflo-setal83a,stenflo-setal83b}. 
This led to the discovery of 
the so-called Second Solar Spectrum, which is the linearly polarized
spectrum that is exclusively caused by coherent scattering processes
on the Sun. For more convenience it is now referred to with the
acronym SS2, implying that the ordinary intensity spectrum is SS1. 

\begin{figure*}
\centering
\includegraphics[width=0.95\textwidth]{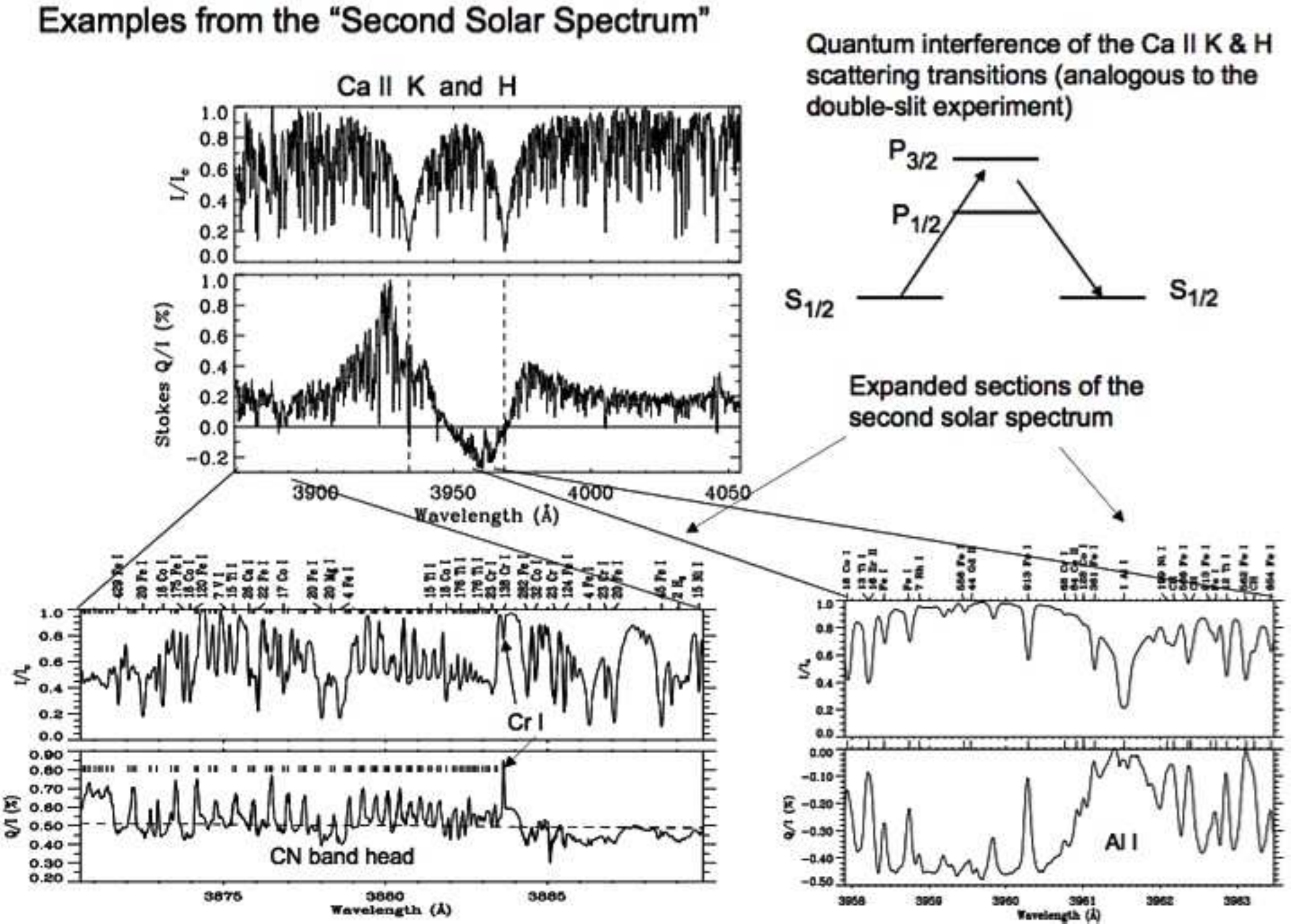}
\caption{Examples of various structures in the Second Solar Spectrum
  (SS2). The panels to the upper left show a nearly 200\,\AA\ wide
  portion around the strong Ca\,{\sc ii} K and H resonance lines
  (marked by the vertical dashed lines). The upper panel shows the
  intensity spectrum (SS1), the panel below SS2 with its remarkable
   polarization sign reversal, which has its explanation in
  terms of quantum interference between the K and H line scattering
  amplitudes, as indicated to the upper right. With the imaging
  polarimeter ZIMPOL it became possible in the late 1990s to zoom in
  on small sections of the spectrum, to uncover the enormous wealth of
  new kinds of structures in SS2, as indicated in the sets of panels
  in the lower half of the figure. }
\label{fig:ss2} \end{figure*}

There is a need to refer to SS2 as a totally different spectrum
because the fractional linear polarization Stokes
$Q/I$ is as richly spectrally structured as the ordinary intensity spectrum, but
with structures that look entirely different and which are formed by largely
different physical processes. Its discovery led to intense theoretical
work trying to identify all the unfamiliar spectral features and 
develop the quantum-mechanical foundations for their interpretations. 
SS2 is a playground for the Hanle effect, but the underlying physics is
complex and still not fully understood. A modern atlas of SS2 has been
produced by
\citet{stenflo-gandorf00,stenflo-gandorf02,stenflo-gandorf05} and is
now available both in pdf and data format at \\ 
http://www.irsol.ch/data\_archive/\#ss2 . 

Figure \ref{fig:ss2} gives an example of SS2 around the
K and H lines of ionized calcium, to indicate the nature of some of
the phenomena that we encounter in SS2. The
nearly 200\,\AA\ wide section shown in the double panel to the upper left was
recorded in 1978 \citep{stenflo-s80}. The upper panel shows SS1 (the
intensity spectrum) with the two dominating Ca\,{\sc ii} resonance
lines, whose extended damping wings form a quasi-continuum that is 
populated by a multitude of blend lines. The strange shape of the SS2
spectrum in the panel below is a signature of quantum-mechanical
interference between the $J=1/2$ and 3/2 upper states of the H and K
lines \citep{stenflo-s80}, as indicated in the diagram to the upper
right. The intermediate state of the coherent scattering 
process represents a ``Schr\"odinger cat state'', a coherent superposition
of states having different total angular momenta. The resulting
quantum interference is responsible for the sign reversal of SS2
in Fig.~\ref{fig:ss2}. 

Although major SS2 features like those of the H and K lines in the
upper part of Fig.~\ref{fig:ss2} could be uncovered with the Kitt Peak
observations in 1978, the polarimetric precision that could be achieved
was not much better than 0.1\,\%, far insufficient for most of the
SS2. It was only with the introduction of the ZIMPOL (Zurich Imaging
Polarimeter) technology in the mid 1990s  that the full wealth of
spectral structures in SS2 became accessible to exploration. With
ZIMPOL at Kitt Peak we could
routinely reach a polarimetric precision better than $10^{-5}$ in
combination with high spectral resolution. The lower
half of Fig.~\ref{fig:ss2} illustrates how this allowed us to zoom
in on any small spectral section to expose the new world of
polarization phenomena. 

\begin{figure*}\sidecaption
\includegraphics[width=0.34\textwidth]{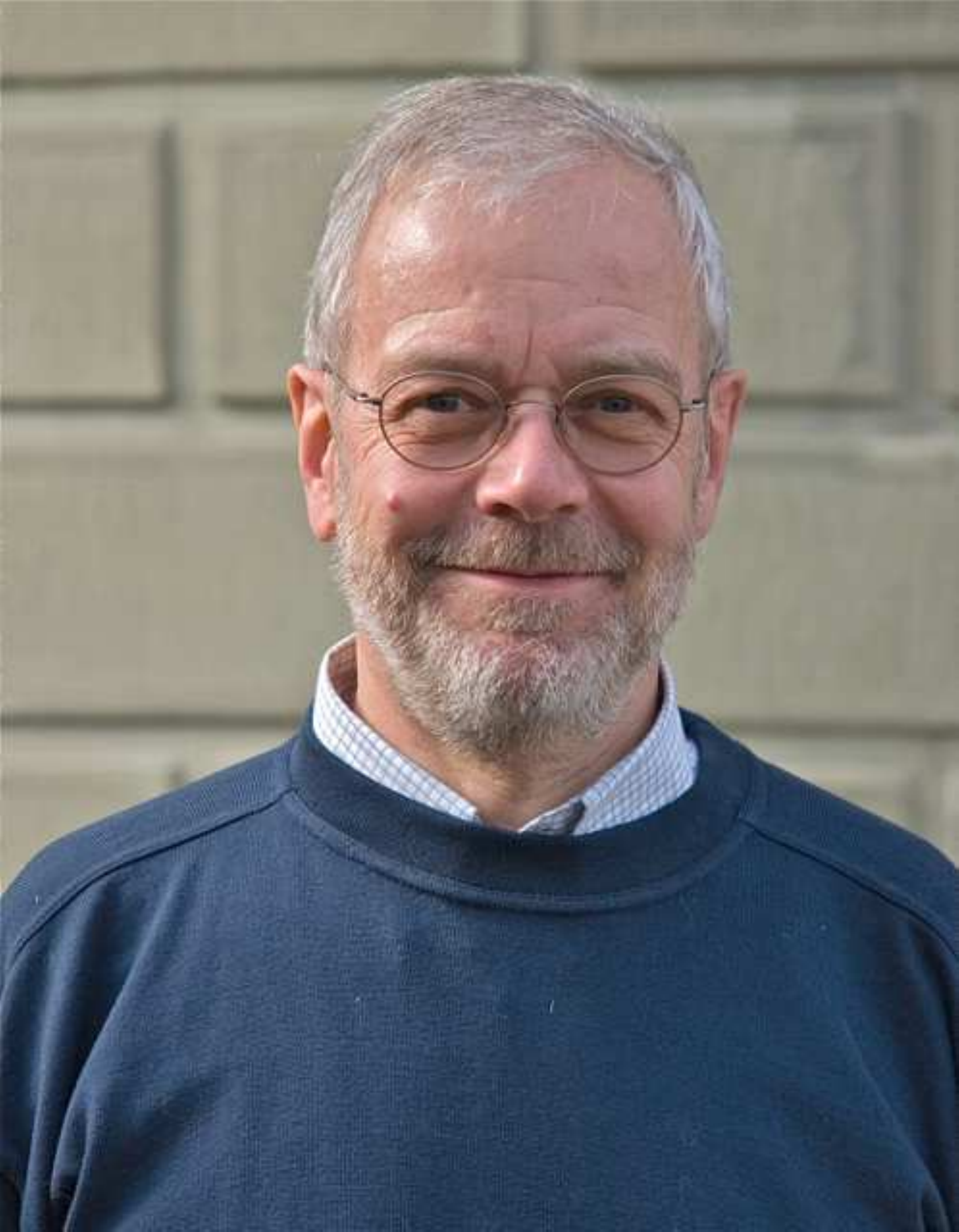}
\caption{Hanspeter Povel, who at ETH Zurich invented the ZIMPOL
  technology for CCD-based imaging polarimetry with a precision of
  better than $10^{-5}$ in the fractional polarization.}
\label{fig:povel} \end{figure*}

The ZIMPOL technology, which was invented by Hanspeter Povel at ETH
Zurich \citep{stenflo-povel95,stenflo-povel01}, revolutionized imaging polarimetry by
achieving a polarimetric precision nearly two orders of magnitude
better than other imaging systems. This opened a window to
polarization physics that had not been accessible before. The
problem had been that the available 2-D electronic detectors of CCD
type have slow readout, which seemed incompatible with the fast (kHz)
polarization modulation needed to eliminate
seeing noise. If instead of modulation one would form the difference between two
simultaneous images in orthogonal polarization states, the accuracy
would be limited by gain-table noise, which is caused by inaccuracies
in the flat-field calibration. Povel's solution to
eliminate these two dominating noise sources was a way to create hidden fast
buffer storage areas within the CCD sensor, between which the
photocharges could be cycled in synchrony with the kHz polarization
modulation. The second and later generations of the ZIMPOL system have
four sets of fast hidden buffers, implying that the photocharges
can be cycled between four simultaneous image planes within a single
CCD sensor. Linear combinations between these four images give us the
simultaneous images of the four Stokes parameters, free from seeing
and gain table noise. The technology has now been successfully used in solar
research for two decades.  

\begin{figure*}\sidecaption
\includegraphics[width=0.60\textwidth]{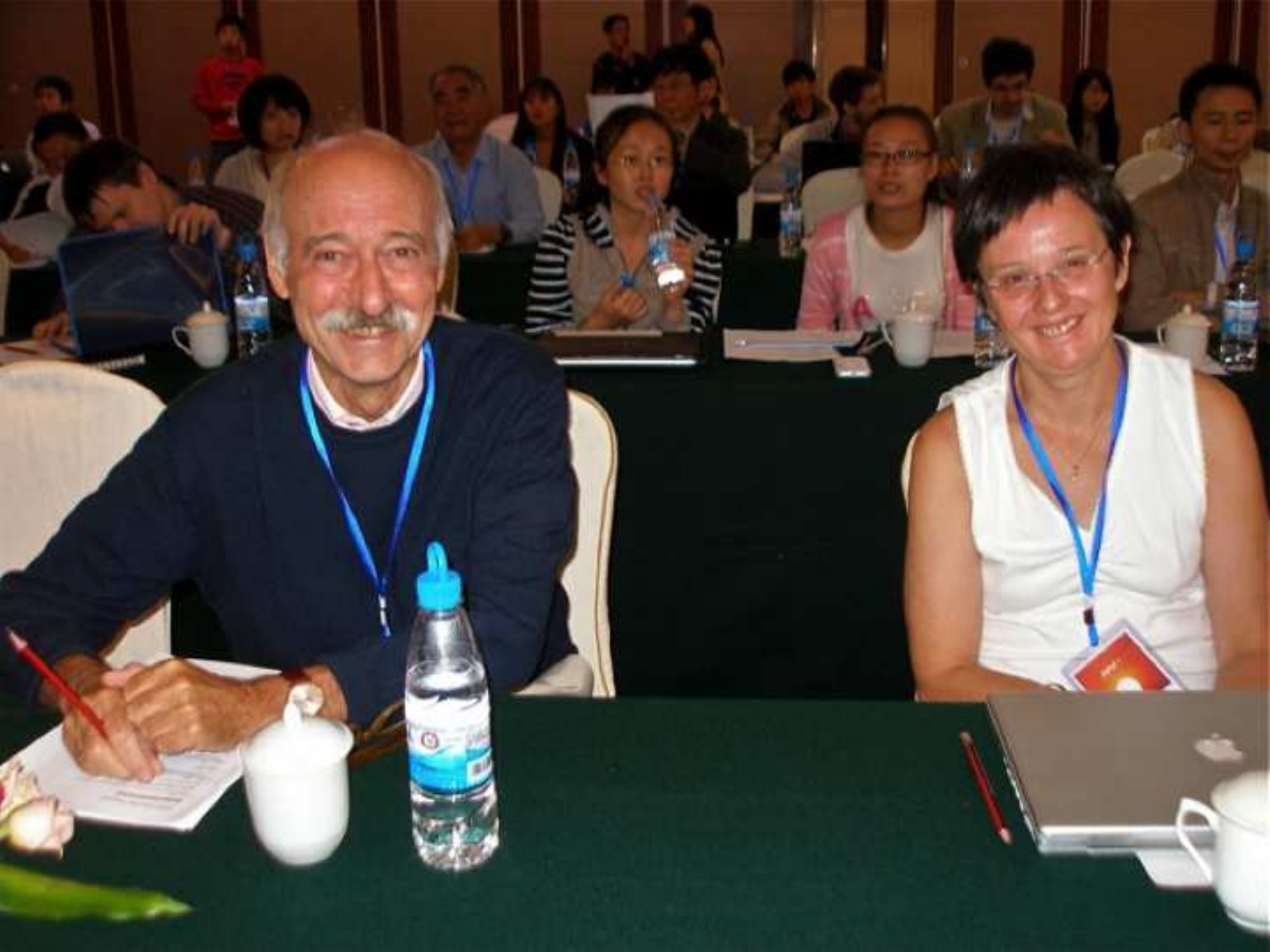}
\caption{Egidio Landi Degl'Innocenti of Firenze and Veronique
  Bommier of Paris in 2013 during Solar Polarization Workshop No.~7 in
Kunming, China. They helped establish the quantum-mechanical
foundation for the interaction between radiation and matter in magnetized media.}
\label{fig:landi} \end{figure*}

The interpretation of the wealth of new phenomena in SS2 and the
application of the Hanle effect for the diagnostics of solar magnetic
fields require that the fundamental quantum processes
that govern the interaction between matter and radiation in magnetized
media are sufficiently understood. Figure \ref{fig:landi} shows two of
the main pioneers in the work to 
establish a foundation based on general principles in
quantum field theory: Egidio Landi Degl'Innocenti of Firenze,
Italy, and Veronique Bommier of Paris. They developed the
theory in 
terms of the density matrix formalism with irreducible tensors,
treated optical pumping by solving the statistical equilibrium problem
for multi-level systems with atomic polarization, and applied the
theory for the diagnostics of prominence magnetic fields 
\citep[cf.][]{stenflo-bommier80,stenflo-landi83}. The
monograph by \citet{stenflo-lanlan04} provides a 
comprehensive account of this theory. 

An alternative approach to the scattering theory in terms of the
Kramers-Heisenberg scattering amplitudes is presented in the monograph
on solar magnetic fields by \citet{stenflo-book94}. It is
particularly suited for the inclusion of the important effects of partial
frequency redistribution effects in polarized
radiative transfer with scattering in magnetized media.  It has been the
approach used by the groups in Bangalore (headed by K.N. Nagendra)
and Nice (led by Marianne Faurobert and Hel\`ene Frisch), who have
developed the theoretical tools needed to interpret the SS2. 

The physics of polarized radiation with the Zeeman and Hanle effects
in magnetized stellar atmospheres has been the subject of a series of
Solar Polarization Workshops that have taken place every three years
in different parts of the world since 1995. The first of them was
organized in St. Petersburg, Russia, by V.V. Ivanov, who coined the
expression ``Second Solar Spectrum'', when referring to the linearly
polarized solar spectrum that is formed by coherent scattering
processes. The proceedings of these Workshops provide an updated
account of the field. The latest is from Solar
Polarization 7, which was held in Kunming in 2013 \citep{stenflo-spw7book} 

%%%%%%%%%%%%%%%%%%%%%%%%%%%%%%%%%%%
\subsection{Scale spectrum of magnetic structures: empirical 
  view}\label{sec:scalesl}
The ongoing quest to diagnose the smallest scales has led to the
understanding that the magnetic structuring continues on scales orders
of magnitude smaller than the resolved ones, down to the magnetic
diffusion limit (of order 25\,m), where the ohmic diffusion time scale
becomes smaller than the convective time scale and the field lines
therefore cease to be frozen-in and decouple from the plasma. Since
most scales are still unresolved, indirect methods based on the
observations of ensemble averages (see Sect.~\ref{sec:ensemble}) have had to be used,
with guidance from numerical simulations of magneto-convection, which
have become increasingly advanced with the rapid growth in computing
power. 

\begin{figure*}\sidecaption
\includegraphics[width=0.65\textwidth]{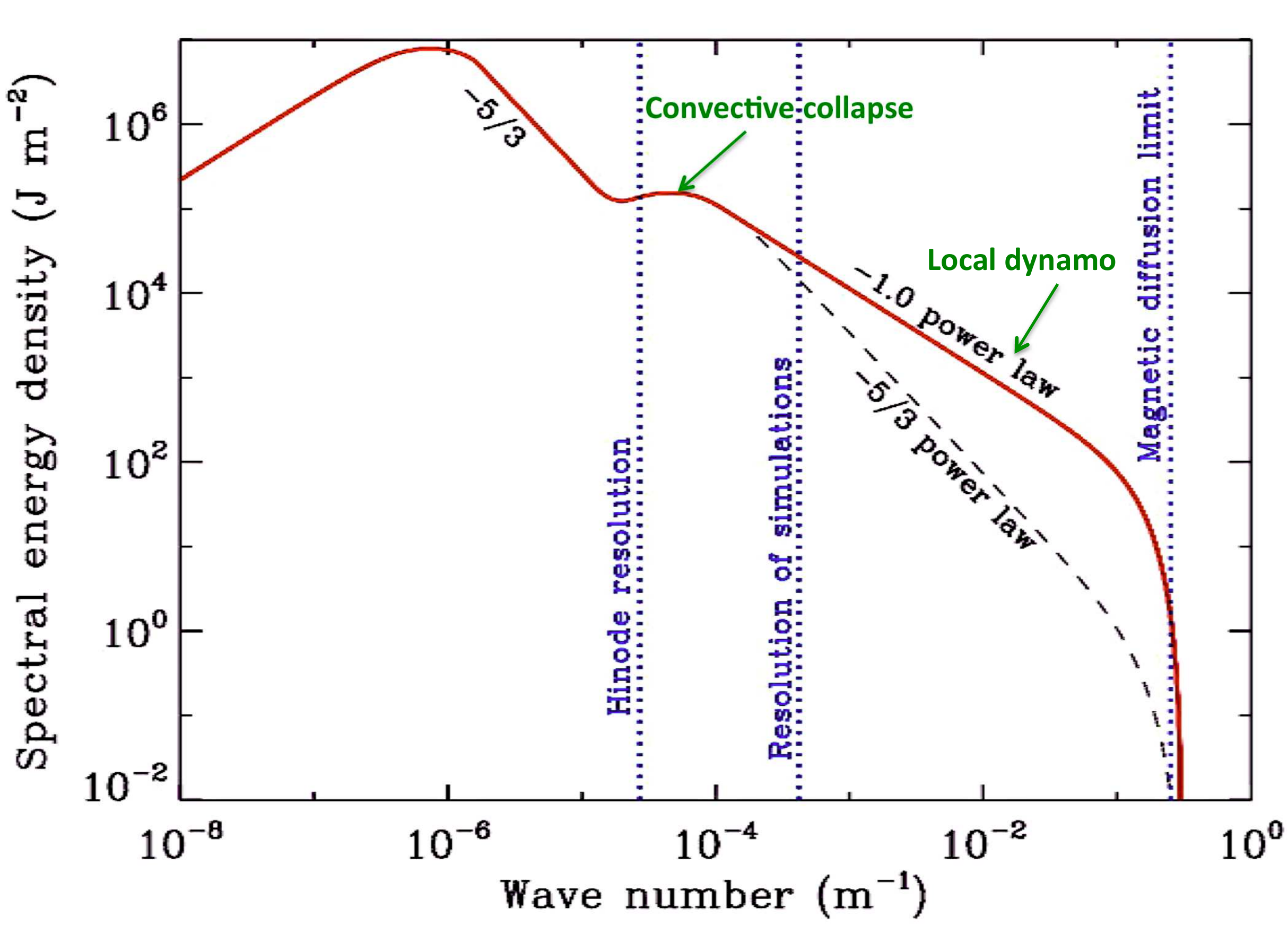}
\caption{Empirically determined magnetic energy spectrum extending
  over seven orders of 
  magnitude in scale (red line), from the global scales to the magnetic diffusion 
limit, where the field lines decouple from the
plasma. The spectrum needs to be elevated in the optically thin
inertial range to a power law of $-1.0$ to be compatible with the
constraints from the observed Hanle depolarization effect. From \citet{stenflo-s14spw7}.}
\label{fig:enspec} \end{figure*}

Figure \ref{fig:enspec} gives a global overview of the empirically
determined magnetic energy spectrum of the quiet Sun over seven orders
of magnitude in scale size, from \citet{stenflo-s14spw7}. The portion
to the left of the vertical 
dotted line marked ``Hinode resolution'' (at a scale of about 200\,km)
is based on resolved observations with {\it Hinode} SOT/SP. The spectral
bump immediately to the right of this line, at scales 10-100\,km,
follows from diagnostics with the line-ratio method, from which a
population of intermittent kG flux elements with sizes in this range
can be inferred. Theoretically the existence of such a flux population
can be understood in terms of the convective collapse mechanism
\citep{stenflo-parker78,stenflo-sprzw79}. 

One of the main applications of the Hanle effect has been to diagnose
the properties of magnetic flux which is hidden to the Zeeman effect,
because the magnetic elements are optically thin with random
orientations of their field vectors. When this method was first
introduced \citep{stenflo-s82} it revealed the existence of a seething
ocean of microturbulent magnetic flux with field strengths in the
range 10-100\,G. The most
detailed radiative-transfer modelling of 
this Hanle depolarization effect with the use of 3-D atmospheres produced
by MHD numerical simulations has led to a turbulent field strength of
60\,G if a single-valued field is assumed ($\delta$ function
PDF). With a more realistic PDF an average field strength about twice
as large is obtained \citep{stenflo-trujetal04}. The 60\,G value
should therefore be considered as a lower limit. 

So much volume-filling flux in the optically thin scale range is not
compatible with the standard Kolmogorov-type $-5/3$ power law (dashed
line at high wave numbers in Fig.~\ref{fig:enspec}). The 
energy spectrum needs to be substantially elevated and 
flatter to be compatible with the Hanle observational
constraints, which are best satisfied with the $-1.0$
power law that is illustrated in Fig.~\ref{fig:enspec}. Such an
elevated spectrum cannot be explained in terms of the ordinary
turbulent cascade from larger to smaller scales, there is a need for a
new source of flux at small scales. This source could be provided by a
local dynamo. While observations at resolved scales rule out a
significant contribution from a local dynamo \citep{stenflo-s12aa2}, such
dynamo action appears to be needed for consistency with the Hanle constraints. 

%%%%%%%%%%%%%%%%%%%%%%%%%%%%%%%
\section{Concluding remarks}\label{sec:concl}
Current research on solar magnetic fields is very diverse, with an
interplay between theory (e.g. plasma physics, quantum field theory,
polarized radiative transfer), observations (ground-based or
space-based, polarimetric imaging with spectrographs or narrow-band
filter systems, etc.), technological developments (detector
and modulation systems, telescope technology, enhanced resolutions,
computing power), and numerical simulations of magneto-convection. The
present rather sketchy and personal overview has focused on some
highlights that have advanced our empirical view of solar magnetic
fields since the time of Hale. It is unavoidable that many of the
important contributions to the field have not been covered here,
because it is far beyond the scope of the present overview to aim
for a comprehensive historical account. 

Hale opened a new window to the universe by applying the physical
effect discovered by Pieter Zeeman a decade before \citep{stenflo-zeeman1897}. A
new polarimetric window has now been opened, although with much delay,
by applying the physical effect discovered by Wilhelm Hanle at the
time when quantum mechanics was being formulated in G\"ottingen
\citep{stenflo-hanle24}.  The complexity of all the coherence
phenomena has led to a return to an exploration of the fundamental
quantum nature of matter-radiation 
interactions in magnetized media. This work is still ongoing. 

In contrast the non-coherent Zeeman-effect physics is well
understood. The problems when applying it to the Sun arise because we
sample an inhomogeneous medium in which the physical parameters vary
over regions that are not resolved. Even if we could observe with
infinite angular resolution, the spatial sampling along the line of
sight remains about 100\,km, since this is the typical 
depth of line formation. This is more than three orders of magnitude
larger than the small-scale end of the magnetic energy
spectrum. One may in principle improve the resolution along the line
of sight by using combinations of spectral lines, whose contribution
functions are slightly shifted relative to each other, but this
improvement can hardly exceed one order of magnitude. Therefore the
observed quantities will always 
represent ensemble averages over distributions of unresolved
structures. The diagnostics of the intrinsic properties of the
elements that make up such ensembles is a formidable challenge. It is
however a challenge that non-solar astronomers are well familiar with,
since for much more distant objects nobody can be misled to believe
that the physical structures are resolved. 

It is therefore obvious that the history of solar magnetic field
observations will not come to an end point any time soon. As much else
in science it is an open-ended enterprise. The past history has
seen many spectacular advances, which have given us insights about the
role played by magnetic fields not only in the Sun but throughout the
universe. The deepened understanding has led to new questions that
could not be contemplated at the time of Hale. The future exploration
of these questions on the Sun will continue to fertilize the rest of
astrophysics. 

%\begin{acknowledgements}
%\end{acknowledgements}

%%%%%%%%%%%%%%%%%%%%%%%%%%%%%


\begin{thebibliography}{72}
% BibTex style file: aps.bst (nameyear), 2011-02-21
\ifx \bisbn   \undefined \def \bisbn  #1{ISBN #1}\fi
\ifx \binits  \undefined \def \binits#1{#1} \fi
\ifx \bauthor  \undefined \def \bauthor#1{#1} \fi
\ifx \bjtitle  \undefined \def \bjtitle#1{\textrm{#1}}\fi
\ifx \batitle  \undefined \def \batitle#1{#1} \fi
\ifx \bctitle  \undefined \def \bctitle#1{#1} \fi
\ifx \bvolume  \undefined \def \bvolume#1{\textbf{#1}}\fi
\ifx \byear  \undefined \def \byear#1{#1} \fi
\ifx \bissue  \undefined \def \bissue#1{#1} \fi
\ifx \bfpage  \undefined \def \bfpage#1{#1} \fi
\ifx \blpage  \undefined \def \blpage #1{#1} \fi
\ifx \burl  \undefined \def \burl#1{#1} \fi
\ifx \doiurl  \undefined \def \doiurl#1{#1} \fi
\ifx \betal  \undefined \def \betal{et al.} \fi
\ifx \binstitute  \undefined \def \binstitute#1{#1} \fi
\ifx \beditor  \undefined \def \beditor#1{#1} \fi
\ifx \bpublisher  \undefined \def \bpublisher#1{#1} \fi
\ifx \bbtitle  \undefined \def \bbtitle#1{\textit{#1}} \fi
\ifx \bedition  \undefined \def \bedition#1{#1} \fi
\ifx \bseriesno  \undefined \def \bseriesno#1{#1} \fi
\ifx \blocation  \undefined \def \blocation#1{#1} \fi
\ifx \bsertitle  \undefined \def \bsertitle#1{#1} \fi
\ifx \bsnm \undefined \def \bsnm#1{#1} \fi
\ifx \bsuffix \undefined \def \bsuffix#1{#1} \fi
\ifx \bparticle \undefined \def \bparticle#1{#1} \fi
\ifx \barticle \undefined \def \barticle#1{#1} \fi
\ifx \botherref \undefined \def \botherref #1{#1} \fi
\ifx \url \undefined \def \url#1{#1} \fi
\ifx \bchapter \undefined \def \bchapter#1{#1} \fi
\ifx \bbook \undefined \def \bbook#1{#1} \fi
\ifx \bcomment \undefined \def \bcomment#1{#1} \fi
\ifx \oauthor \undefined \def \oauthor#1{#1} \fi
\ifx \citeauthoryear \undefined \def \citeauthoryear#1{#1} \fi
\ifx \texttildelow  \undefined \def \texttildelow{\symbol{126}} \fi
\def \endbibitem {}
\ifx \bconflocation  \undefined \def \bconflocation#1{#1} \fi

\bibitem[\protect\citeauthoryear{{Alfv{\'e}n}}{1942}]{stenflo-alfven42}
\begin{barticle}
\bauthor{\binits{H.} \bsnm{{Alfv{\'e}n}}},
\batitle{{Existence of Electromagnetic-Hydrodynamic Waves}}.
\bjtitle{\nat}
\bvolume{150},
\bfpage{405}--\blpage{406}
(\byear{1942})
\end{barticle}
\endbibitem

\bibitem[\protect\citeauthoryear{{Alfv{\'e}n}}{1967}]{stenflo-alfven67}
\begin{bchapter}
\bauthor{\binits{H.} \bsnm{{Alfv{\'e}n}}},
\bctitle{{Solar Magnetic Fields}},
in \bbtitle{Magnetism and the Cosmos},
ed. by \beditor{\binits{W.R.} \bsnm{{Hindmarsh}}},
\beditor{\binits{F.J.} \bsnm{{Lowes}}},
\beditor{\binits{P.H.} \bsnm{{Roberts}}},
\beditor{\binits{F.R.} \bsnm{{Runcorn}}},
(\bpublisher{Oliver \&\ Boyd}, \blocation{Edinburgh}, 
\byear{1967}),
pp. \bfpage{246}--\blpage{261}
\end{bchapter}
\endbibitem

\bibitem[\protect\citeauthoryear{{Alfv{\'e}n} and
  {Carlqvist}}{1967}]{stenflo-alfvencarlqvist67}
\begin{barticle}
\bauthor{\binits{H.} \bsnm{{Alfv{\'e}n}}},
\bauthor{\binits{P.} \bsnm{{Carlqvist}}},
\batitle{{Currents in the Solar Atmosphere and a Theory of Solar Flares}}.
\bjtitle{\solphys}
\bvolume{1},
\bfpage{220}--\blpage{228}
(\byear{1967})
\end{barticle}
\endbibitem

\bibitem[\protect\citeauthoryear{{Babcock}}{1953}]{stenflo-babcock53}
\begin{barticle}
\bauthor{\binits{H.W.} \bsnm{{Babcock}}},
\batitle{{The Solar Magnetograph}}.
\bjtitle{\apj}
\bvolume{118},
\bfpage{387}
(\byear{1953})
\end{barticle}
\endbibitem

\bibitem[\protect\citeauthoryear{{Babcock}}{1961}]{stenflo-babcock61}
\begin{barticle}
\bauthor{\binits{H.W.} \bsnm{{Babcock}}},
\batitle{{The Topology of the Sun's Magnetic Field and the 22-Year Cycle.}}
\bjtitle{\apj}
\bvolume{133},
\bfpage{572}--\blpage{587}
(\byear{1961})
\end{barticle}
\endbibitem

\bibitem[\protect\citeauthoryear{{Bommier}}{1980}]{stenflo-bommier80}
\begin{barticle}
\bauthor{\binits{V.} \bsnm{{Bommier}}},
\batitle{{Quantum theory of the Hanle effect. II - Effect of level-crossings
  and anti-level-crossings on the polarization of the D3 helium line of solar
  prominences}}.
\bjtitle{\aap}
\bvolume{87},
\bfpage{109}--\blpage{120}
(\byear{1980})
\end{barticle}
\endbibitem

\bibitem[\protect\citeauthoryear{{Brault}}{1978}]{stenflo-brault78}
\begin{barticle}
\bauthor{\binits{J.W.} \bsnm{{Brault}}},
\batitle{{Solar Fourier Transform Spectroscopy}}.
\bjtitle{Osserv. Mem. Oss. Astrofis. Arcetri}
\bvolume{106},
\bfpage{33}
(\byear{1978})
\end{barticle}
\endbibitem

\bibitem[\protect\citeauthoryear{{Brault}}{1985}]{stenflo-brault85}
\begin{bchapter}
\bauthor{\binits{J.W.} \bsnm{{Brault}}},
\bctitle{{Fourier Transform Spectroscopy}},
in \bbtitle{High Resolution in Astronomy. Fifteenth Advanced Course of the
  Swiss Society of Astronomy and Astrophysics},
ed. by \beditor{\binits{A.O.} \bsnm{{Benz}}},
\beditor{\binits{M.} \bsnm{{Huber}}},
\beditor{\binits{M.} \bsnm{{Mayor}}}
(\bpublisher{Geneva Obs.}, \blocation{Sauverny}, \byear{1985}),
pp. \bfpage{3}--\blpage{61}
\end{bchapter}
\endbibitem

\bibitem[\protect\citeauthoryear{{Br{\"u}ckner}}{1963}]{stenflo-brueckner63}
\begin{barticle}
\bauthor{\binits{G.} \bsnm{{Br{\"u}ckner}}},
\batitle{{Photoelektrische Polarisationsmessungen an Resonanzlinien im
  Sonnenspektrum. Mit 6 Textabbildungen}}.
\bjtitle{\zap}
\bvolume{58},
\bfpage{73}
(\byear{1963})
\end{barticle}
\endbibitem

\bibitem[\protect\citeauthoryear{{Chapman} and
  {Sheeley}}{1968}]{stenflo-chsh68}
\begin{barticle}
\bauthor{\binits{G.A.} \bsnm{{Chapman}}},
\bauthor{\binits{N.R.} \bsnm{{Sheeley}} \bsuffix{Jr.}},
\batitle{{The Photospheric Network}}.
\bjtitle{\solphys}
\bvolume{5},
\bfpage{442}--\blpage{461}
(\byear{1968})
\end{barticle}
\endbibitem

\bibitem[\protect\citeauthoryear{{Cowling}}{1933}]{stenflo-cowling33}
\begin{barticle}
\bauthor{\binits{T.G.} \bsnm{{Cowling}}},
\batitle{{The magnetic field of sunspots}}.
\bjtitle{\mnras}
\bvolume{94},
\bfpage{39}--\blpage{48}
(\byear{1933})
\end{barticle}
\endbibitem

\bibitem[\protect\citeauthoryear{{Dumont} et~al.}{1973}]{stenflo-dumontetal73}
\begin{barticle}
\bauthor{\binits{S.} \bsnm{{Dumont}}},
\bauthor{\binits{J.-C.} \bsnm{{Pecker}}},
\bauthor{\binits{A.} \bsnm{{Omont}}},
\batitle{{Theoretical study of the Fraunhofer lines polarization: The case of
  Ca i 4227}}.
\bjtitle{\solphys}
\bvolume{28},
\bfpage{271}--\blpage{288}
(\byear{1973})
\end{barticle}
\endbibitem

\bibitem[\protect\citeauthoryear{{Elsasser}}{1946}]{stenflo-elsasser1946}
\begin{barticle}
\bauthor{\binits{W.M.} \bsnm{{Elsasser}}},
\batitle{{Induction Effects in Terrestrial Magnetism Part I. Theory}}.
\bjtitle{Phys. Rev.}
\bvolume{69},
\bfpage{106}--\blpage{116}
(\byear{1946})
\end{barticle}
\endbibitem

\bibitem[\protect\citeauthoryear{{Elsasser}}{1955}]{stenflo-elsasser1955}
\begin{barticle}
\bauthor{\binits{W.M.} \bsnm{{Elsasser}}},
\batitle{{Hydromagnetism. I. A Review}}.
\bjtitle{Amer. J. Phys.}
\bvolume{23},
\bfpage{590}--\blpage{609}
(\byear{1955})
\end{barticle}
\endbibitem

\bibitem[\protect\citeauthoryear{{Engvold}}{1991}]{stenflo-engvold91}
\begin{barticle}
\bauthor{\binits{O.} \bsnm{{Engvold}}},
\batitle{{Large Earth-based Solar Telescope - LEST}}.
\bjtitle{Advances in Space Research}
\bvolume{11},
\bfpage{157}--\blpage{168}
(\byear{1991})
\end{barticle}
\endbibitem

\bibitem[\protect\citeauthoryear{{Gandorfer}}{2000}]{stenflo-gandorf00}
\begin{bbook}
\bauthor{\binits{A.} \bsnm{{Gandorfer}}},
\bbtitle{{The Second Solar Spectrum: A High Spectral Resolution Polarimetric
  Survey of Scattering Polarization at the Solar Limb in Graphical
  Representation. Volume I: 4625 {\rm \AA} to 6995 {\rm \AA}}}
(\bpublisher{VdF}, \blocation{Zurich}, \byear{2000})
\end{bbook}
\endbibitem

\bibitem[\protect\citeauthoryear{{Gandorfer}}{2002}]{stenflo-gandorf02}
\begin{bbook}
\bauthor{\binits{A.} \bsnm{{Gandorfer}}},
\bbtitle{{The Second Solar Spectrum: A High Spectral Resolution Polarimetric
  Survey of Scattering Polarization at the Solar Limb in Graphical
  Representation. Volume II: 3910 {\rm \AA} to 4630 {\rm \AA}}}
(\bpublisher{VdF}, \blocation{Zurich}, \byear{2002})
\end{bbook}
\endbibitem

\bibitem[\protect\citeauthoryear{{Gandorfer}}{2005}]{stenflo-gandorf05}
\begin{bbook}
\bauthor{\binits{A.} \bsnm{{Gandorfer}}},
\bbtitle{{The Second Solar Spectrum: A High Spectral Resolution Polarimetric
  Survey of Scattering Polarization at the Solar Limb in Graphical
  Representation. Volume III: 3160 {\rm \AA} to 3915 {\rm \AA}}}
(\bpublisher{VdF}, \blocation{zurich}, \byear{2005})
\end{bbook}
\endbibitem

\bibitem[\protect\citeauthoryear{{Hale}}{1908}]{stenflo-hale08}
\begin{barticle}
\bauthor{\binits{G.E.} \bsnm{{Hale}}},
\batitle{{On the Probable Existence of a Magnetic Field in Sun-Spots}}.
\bjtitle{\apj}
\bvolume{28},
\bfpage{315}
(\byear{1908})
\end{barticle}
\endbibitem

\bibitem[\protect\citeauthoryear{{Hale} et~al.}{1918}]{stenflo-haleetal18}
\begin{barticle}
\bauthor{\binits{G.E.} \bsnm{{Hale}}},
\bauthor{\binits{F.H.} \bsnm{{Seares}}},
\bauthor{\binits{A.} \bsnm{{van Maanen}}},
\bauthor{\binits{F.} \bsnm{{Ellerman}}},
\batitle{{The General Magnetic Field of the Sun. Apparent Variation of
  Field-Strength with Level in the Solar Atmosphere}}.
\bjtitle{\apj}
\bvolume{47},
\bfpage{206}--\blpage{254}
(\byear{1918})
\end{barticle}
\endbibitem

\bibitem[\protect\citeauthoryear{{Hale} et~al.}{1919}]{stenflo-haleetal19}
\begin{barticle}
\bauthor{\binits{G.E.} \bsnm{{Hale}}},
\bauthor{\binits{F.} \bsnm{{Ellerman}}},
\bauthor{\binits{S.B.} \bsnm{{Nicholson}}},
\bauthor{\binits{A.H.} \bsnm{{Joy}}},
\batitle{{The Magnetic Polarity of Sun-Spots}}.
\bjtitle{\apj}
\bvolume{49},
\bfpage{153}--\blpage{178}
(\byear{1919})
\end{barticle}
\endbibitem

\bibitem[\protect\citeauthoryear{{Hanle}}{1924}]{stenflo-hanle24}
\begin{barticle}
\bauthor{\binits{W.} \bsnm{{Hanle}}},
\batitle{{\"Uber magnetische Beeinflussung der Polarisation der
  Resonanzfluoreszenz}}.
\bjtitle{Z. Phys.}
\bvolume{30},
\bfpage{93}--\blpage{105}
(\byear{1924})
\end{barticle}
\endbibitem

\bibitem[\protect\citeauthoryear{{Harvey}}{1977}]{stenflo-harvey77}
\begin{barticle}
\bauthor{\binits{J.W.} \bsnm{{Harvey}}},
\batitle{{Observations of Small-Scale Photospheric Magnetic Fields}}.
\bjtitle{Highlights of Astronomy}
\bvolume{4},
\bfpage{223}
(\byear{1977})
\end{barticle}
\endbibitem

\bibitem[\protect\citeauthoryear{{Harvey} and
  {Martin}}{1973}]{stenflo-harveymartin73}
\begin{barticle}
\bauthor{\binits{K.L.} \bsnm{{Harvey}}},
\bauthor{\binits{S.F.} \bsnm{{Martin}}},
\batitle{{Ephemeral Active Regions}}.
\bjtitle{\solphys}
\bvolume{32},
\bfpage{389}--\blpage{402}
(\byear{1973})
\end{barticle}
\endbibitem

\bibitem[\protect\citeauthoryear{{Ishikawa} et~al.}{2011}]{stenflo-claspspw6}
\begin{bchapter}
\bauthor{\binits{R.} \bsnm{{Ishikawa}}},
\bauthor{\binits{T.} \bsnm{{Bando}}},
\bauthor{\binits{D.} \bsnm{{Fujimura}}},
\bauthor{\binits{H.} \bsnm{{Hara}}},
\bauthor{\binits{R.} \bsnm{{Kano}}},
\bauthor{\binits{T.} \bsnm{{Kobiki}}},
\bauthor{\binits{N.} \bsnm{{Narukage}}},
\bauthor{\binits{S.} \bsnm{{Tsuneta}}},
\bauthor{\binits{K.} \bsnm{{Ueda}}},
\bauthor{\binits{H.} \bsnm{{Wantanabe}}},
\bauthor{\binits{K.} \bsnm{{Kobayashi}}},
\bauthor{\binits{J.} \bsnm{{Trujillo Bueno}}},
\bauthor{\binits{R.} \bsnm{{Manso Sainz}}},
\bauthor{\binits{J.} \bsnm{{Stepan}}},
\bauthor{\binits{B.} \bsnm{{de Pontieu}}},
\bauthor{\binits{M.} \bsnm{{Carlsson}}},
\bauthor{\binits{R.} \bsnm{{Casini}}},
\bctitle{{A Sounding Rocket Experiment for Spectropolarimetric Observations
  with the Ly$_\alpha$ Line at 121.6 nm (CLASP)}},
in \bbtitle{Solar Polarization 6},
ed. by \beditor{\binits{J.R.} \bsnm{{Kuhn}}},
\beditor{\binits{D.M.} \bsnm{{Harrington}}},
\beditor{\binits{H.} \bsnm{{Lin}}},
\beditor{\binits{S.V.} \bsnm{{Berdyugina}}},
\beditor{\binits{J.} \bsnm{{Trujillo-Bueno}}},
\beditor{\binits{S.L.} \bsnm{{Keil}}},
\beditor{\binits{T.} \bsnm{{Rimmele}}}. 
\bsertitle{Astronomical Society of the Pacific Conference Series} 
\bvolume{437},
\bfpage{287}--\blpage{295}
(\byear{2011})
\end{bchapter}
\endbibitem

\bibitem[\protect\citeauthoryear{{Kiepenheuer}}{1973}]{stenflo-kiepenheuer73}
\begin{bchapter}
\bauthor{\binits{K.O.} \bsnm{{Kiepenheuer}}},
\bctitle{{Joint Organization for Solar Observations (JOSO)}},
in \bbtitle{Solar Activity and Related Interplanetary and Terrestrial
  Phenomena}.  Proceedings of the First European Astronomical Meeting, 
ed. by \beditor{\binits{J.} \bsnm{{Xanthakis}}} 
(\bpublisher{Springer}, \blocation{Heidelberg}, \byear{1973}),
p. \bfpage{165}
\end{bchapter}
\endbibitem

\bibitem[\protect\citeauthoryear{{Landi
  Degl'Innocenti}}{1983}]{stenflo-landi83}
\begin{barticle}
\bauthor{\binits{E.} \bsnm{{Landi Degl'Innocenti}}},
\batitle{{Polarization in spectral lines. I - A unifying theoretical
  approach.}}
\bjtitle{\solphys}
\bvolume{85},
\bfpage{3}--\blpage{31}
(\byear{1983})
\end{barticle}
\endbibitem

\bibitem[\protect\citeauthoryear{{Landi Degl'Innocenti} and
  {Landolfi}}{2004}]{stenflo-lanlan04}
\begin{bbook}
\bauthor{\binits{E.} \bsnm{{Landi Degl'Innocenti}}},
\bauthor{\binits{M.} \bsnm{{Landolfi}}},
\bbtitle{{Polarization in Spectral Lines}}.
\bsertitle{Astrophysics and Space Science Library},
vol. \bseriesno{307}
(\bpublisher{Kluwer}, \blocation{Dordrecht}, \byear{2004})
\end{bbook}
\endbibitem

\bibitem[\protect\citeauthoryear{{Larmor}}{1919}]{stenflo-larmor1919}
\begin{barticle}
\bauthor{\binits{J.} \bsnm{{Larmor}}},
\batitle{{How could a rotating body such as the sun become a magnet?}}
\bjtitle{Rep. Brit. Assoc. Adv. Sci.}
\bvolume{87},
\bfpage{159}--\blpage{160}
(\byear{1919})
\end{barticle}
\endbibitem

\bibitem[\protect\citeauthoryear{{Leighton}}{1959}]{stenflo-leighton59}
\begin{barticle}
\bauthor{\binits{R.B.} \bsnm{{Leighton}}},
\batitle{{Observations of Solar Magnetic Fields in Plage Regions.}}
\bjtitle{\apj}
\bvolume{130},
\bfpage{366}
(\byear{1959})
\end{barticle}
\endbibitem

\bibitem[\protect\citeauthoryear{{Leighton}}{1969}]{stenflo-leighton69}
\begin{barticle}
\bauthor{\binits{R.B.} \bsnm{{Leighton}}},
\batitle{{A Magneto-Kinematic Model of the Solar Cycle}}.
\bjtitle{\apj}
\bvolume{156},
\bfpage{1}--\blpage{26}
(\byear{1969})
\end{barticle}
\endbibitem

\bibitem[\protect\citeauthoryear{{Leighton}
  et~al.}{1962}]{stenflo-leightonetal62}
\begin{barticle}
\bauthor{\binits{R.B.} \bsnm{{Leighton}}},
\bauthor{\binits{R.W.} \bsnm{{Noyes}}},
\bauthor{\binits{G.W.} \bsnm{{Simon}}},
\batitle{{Velocity Fields in the Solar Atmosphere. I. Preliminary Report.}}
\bjtitle{\apj}
\bvolume{135},
\bfpage{474}
(\byear{1962})
\end{barticle}
\endbibitem

\bibitem[\protect\citeauthoryear{{Leroy} et~al.}{1977}]{stenflo-leroy77}
\begin{barticle}
\bauthor{\binits{J.L.} \bsnm{{Leroy}}},
\bauthor{\binits{G.} \bsnm{{Ratier}}},
\bauthor{\binits{V.} \bsnm{{Bommier}}},
\batitle{{The polarization of the D3 emission line in prominences}}.
\bjtitle{\aap}
\bvolume{54},
\bfpage{811}--\blpage{816}
(\byear{1977})
\end{barticle}
\endbibitem

\bibitem[\protect\citeauthoryear{{Livingston}
  et~al.}{1976}]{stenflo-livingstonetal76}
\begin{barticle}
\bauthor{\binits{W.C.} \bsnm{{Livingston}}},
\bauthor{\binits{J.} \bsnm{{Harvey}}},
\bauthor{\binits{C.} \bsnm{{Slaughter}}},
\bauthor{\binits{D.} \bsnm{{Trumbo}}},
\batitle{{Solar magnetograph employing integrated diode arrays}}.
\bjtitle{\ao}
\bvolume{15},
\bfpage{40}--\blpage{52}
(\byear{1976})
\end{barticle}
\endbibitem

\bibitem[\protect\citeauthoryear{{Lundmark}}{1926}]{stenflo-lundmark26}
\begin{barticle}
\bauthor{\binits{K.} \bsnm{{Lundmark}}},
\batitle{{Internal motions of Messier 33.}}
\bjtitle{\apj}
\bvolume{63},
\bfpage{67}--\blpage{71}
(\byear{1926})
\end{barticle}
\endbibitem

\bibitem[\protect\citeauthoryear{{Moruzzi} and {Strumia}}{1991}]{stenflo-mor91}
\begin{bbook}
\beditor{\binits{G.} \bsnm{{Moruzzi}}},
\beditor{\binits{F.} \bsnm{{Strumia}}} (eds.),
\bbtitle{The Hanle Effect and Level-Crossing Spectroscopy.}
(\bpublisher{Plenum Press}, \blocation{New York}, \byear{1991})
\end{bbook}
\endbibitem

\bibitem[\protect\citeauthoryear{{Nagendra} et~al.}{2014}]{stenflo-spw7book}
\begin{bbook}
\beditor{\binits{K.N.} \bsnm{{Nagendra}}},
\beditor{\binits{J.O.} \bsnm{{Stenflo}}},
\beditor{\binits{Z.Q.} \bsnm{{Qu}}},
\beditor{\binits{M.} \bsnm{{Sampoorna}}} (eds.),
\bbtitle{Solar Polarization 7}
\bsertitle{Astronomical Society of the Pacific Conference Series} 
\bvolume{489}
(\byear{2014})
\end{bbook}
\endbibitem

\bibitem[\protect\citeauthoryear{{{\"O}hman}}{1929}]{stenflo-ohman29}
\begin{barticle}
\bauthor{\binits{Y.} \bsnm{{{\"O}hman}}},
\batitle{{Astronomical Consequences of the Polarization of Fluorescence}}.
\bjtitle{\mnras}
\bvolume{89},
\bfpage{479}--\blpage{482}
(\byear{1929})
\end{barticle}
\endbibitem

\bibitem[\protect\citeauthoryear{{Parker}}{1955}]{stenflo-parker55}
\begin{barticle}
\bauthor{\binits{E.N.} \bsnm{{Parker}}},
\batitle{{Hydromagnetic Dynamo Models.}}
\bjtitle{\apj}
\bvolume{122},
\bfpage{293}
(\byear{1955})
\end{barticle}
\endbibitem

\bibitem[\protect\citeauthoryear{{Parker}}{1978}]{stenflo-parker78}
\begin{barticle}
\bauthor{\binits{E.N.} \bsnm{{Parker}}},
\batitle{{Hydraulic concentration of magnetic fields in the solar photosphere.
  VI - Adiabatic cooling and concentration in downdrafts}}.
\bjtitle{\apj}
\bvolume{221},
\bfpage{368}--\blpage{377}
(\byear{1978})
\end{barticle}
\endbibitem

\bibitem[\protect\citeauthoryear{{Povel}}{1995}]{stenflo-povel95}
\begin{barticle}
\bauthor{\binits{H.} \bsnm{{Povel}}},
\batitle{{Imaging Stokes polarimetry with piezoelastic modulators and
  charge-coupled-device image sensors.}}
\bjtitle{Optical Engineering}
\bvolume{34},
\bfpage{1870}--\blpage{1878}
(\byear{1995})
\end{barticle}
\endbibitem

\bibitem[\protect\citeauthoryear{{Povel}}{2001}]{stenflo-povel01}
\begin{bchapter}
\bauthor{\binits{H.P.} \bsnm{{Povel}}},
\bctitle{{Ground-based Instrumentation for Solar Magnetic Field Studies, with
  Special Emphasis on the Zurich Imaging Polarimeters Zimpol-i and Ii}},
in \bbtitle{Magnetic Fields Across the Hertzsprung-Russell Diagram},
ed. by \beditor{\binits{G.} \bsnm{{Mathys}}},
\beditor{\binits{S.K.} \bsnm{{Solanki}}},
\beditor{\binits{D.T.} \bsnm{{Wickramasinghe}}}
\bsertitle{Astronomical Society of the Pacific Conference Series} 
\bvolume{248},
\bfpage{543}--\blpage{552}
(\byear{2001})
\end{bchapter}
\endbibitem

\bibitem[\protect\citeauthoryear{{Rachkovsky}}{1962}]{stenflo-rachkovsky62a}
\begin{barticle}
\bauthor{\binits{D.N.} \bsnm{{Rachkovsky}}},
\batitle{{Magneto-optical effects in spectral lines of sunspots}}.
\bjtitle{Izv. Krymsk. Astrofiz. Obs.}
\bvolume{27},
\bfpage{148}--\blpage{161}
(\byear{1962a})
\end{barticle}
\endbibitem

\bibitem[\protect\citeauthoryear{{Rachkovsky}}{1962}]{stenflo-rachkovsky62b}
\begin{barticle}
\bauthor{\binits{D.N.} \bsnm{{Rachkovsky}}},
\batitle{{Magnetic rotation effects in spectral lines}}.
\bjtitle{Izv. Krymsk. Astrofiz. Obs.}
\bvolume{28},
\bfpage{259}--\blpage{270}
(\byear{1962b})
\end{barticle}
\endbibitem

\bibitem[\protect\citeauthoryear{{Severny}}{1964}]{stenflo-severny64}
\begin{barticle}
\bauthor{\binits{A.} \bsnm{{Severny}}},
\batitle{{Solar Magnetic Fields}}.
\bjtitle{\ssr}
\bvolume{3},
\bfpage{451}--\blpage{486}
(\byear{1964})
\end{barticle}
\endbibitem

\bibitem[\protect\citeauthoryear{{Severny}}{1966}]{stenflo-severny66}
\begin{barticle}
\bauthor{\binits{A.B.} \bsnm{{Severny}}},
\batitle{{Magnetic fields of the Sun and stars (in Russian)}}.
\bjtitle{Uspechi fiz. nauk}
\bvolume{88},
\bfpage{3}--\blpage{50}
(\byear{1966})
\end{barticle}
\endbibitem

\bibitem[\protect\citeauthoryear{{Sheeley}}{1967}]{stenflo-sh67}
\begin{barticle}
\bauthor{\binits{N.R.} \bsnm{{Sheeley}} \bsuffix{Jr.}},
\batitle{{Observations of Small-Scale Solar Magnetic Fields}}.
\bjtitle{\solphys}
\bvolume{1},
\bfpage{171}--\blpage{179}
(\byear{1967})
\end{barticle}
\endbibitem

\bibitem[\protect\citeauthoryear{{Sheeley} et~al.}{1985}]{stenflo-sheeley85}
\begin{barticle}
\bauthor{\binits{N.R.} \bsnm{{Sheeley}} \bsuffix{Jr.}},
\bauthor{\binits{C.R.} \bsnm{{DeVore}}},
\bauthor{\binits{J.P.} \bsnm{{Boris}}},
\batitle{{Simulations of the mean solar magnetic field during sunspot cycle
  21}}.
\bjtitle{\solphys}
\bvolume{98},
\bfpage{219}--\blpage{239}
(\byear{1985})
\end{barticle}
\endbibitem

\bibitem[\protect\citeauthoryear{{Solanki} et~al.}{2010}]{stenflo-sunrise2010}
\begin{barticle}
\bauthor{\binits{S.K.} \bsnm{{Solanki}}},
\bauthor{\binits{P.} \bsnm{{Barthol}}},
\bauthor{\binits{S.} \bsnm{{Danilovic}}},
\bauthor{\binits{A.} \bsnm{{Feller}}},
\bauthor{\binits{A.} \bsnm{{Gandorfer}}},
\bauthor{\binits{J.} \bsnm{{Hirzberger}}},
\bauthor{\binits{T.L.} \bsnm{{Riethm{\"u}ller}}},
\bauthor{\binits{M.} \bsnm{{Sch{\"u}ssler}}},
\bauthor{\binits{J.A.} \bsnm{{Bonet}}},
\bauthor{\binits{V.} \bsnm{{Mart{\'{\i}}nez Pillet}}},
\bauthor{\binits{J.C.} \bsnm{{del Toro Iniesta}}},
\bauthor{\binits{V.} \bsnm{{Domingo}}},
\bauthor{\binits{J.} \bsnm{{Palacios}}},
\bauthor{\binits{M.} \bsnm{{Kn{\"o}lker}}},
\bauthor{\binits{N.} \bsnm{{Bello Gonz{\'a}lez}}},
\bauthor{\binits{T.} \bsnm{{Berkefeld}}},
\bauthor{\binits{M.} \bsnm{{Franz}}},
\bauthor{\binits{W.} \bsnm{{Schmidt}}},
\bauthor{\binits{A.M.} \bsnm{{Title}}},
\batitle{{SUNRISE: Instrument, Mission, Data, and First Results}}.
\bjtitle{\apjl}
\bvolume{723},
\bfpage{127}--\blpage{133}
(\byear{2010})
\end{barticle}
\endbibitem

\bibitem[\protect\citeauthoryear{{Spruit} and
  {Zweibel}}{1979}]{stenflo-sprzw79}
\begin{barticle}
\bauthor{\binits{H.C.} \bsnm{{Spruit}}},
\bauthor{\binits{E.G.} \bsnm{{Zweibel}}},
\batitle{{Convective instability of thin flux tubes}}.
\bjtitle{\solphys}
\bvolume{62},
\bfpage{15}--\blpage{22}
(\byear{1979})
\end{barticle}
\endbibitem

\bibitem[\protect\citeauthoryear{{Steenbeck} and
  {Krause}}{1969}]{stenflo-steenkrause69}
\begin{barticle}
\bauthor{\binits{M.} \bsnm{{Steenbeck}}},
\bauthor{\binits{F.} \bsnm{{Krause}}},
\batitle{{Zur Dynamotheorie stellarer und planetarer Magnetfelder I. Berechnung
  sonnen\"ahnlicher Wechselfeldgeneratoren}}.
\bjtitle{Astron. Nachr.}
\bvolume{291},
\bfpage{49}--\blpage{84}
(\byear{1969})
\end{barticle}
\endbibitem

\bibitem[\protect\citeauthoryear{{Stenflo}}{1970}]{stenflo-s70hale}
\begin{barticle}
\bauthor{\binits{J.O.} \bsnm{{Stenflo}}},
\batitle{{Hale's Attempts to Determine the Sun's General Magnetic Field}}.
\bjtitle{\solphys}
\bvolume{14},
\bfpage{263}--\blpage{273}
(\byear{1970})
\end{barticle}
\endbibitem

\bibitem[\protect\citeauthoryear{{Stenflo}}{1973}]{stenflo-s73}
\begin{barticle}
\bauthor{\binits{J.O.} \bsnm{{Stenflo}}},
\batitle{{Magnetic-Field Structure of the Photospheric Network}}.
\bjtitle{\solphys}
\bvolume{32},
\bfpage{41}--\blpage{63}
(\byear{1973})
\end{barticle}
\endbibitem

\bibitem[\protect\citeauthoryear{{Stenflo}}{1980}]{stenflo-s80}
\begin{barticle}
\bauthor{\binits{J.O.} \bsnm{{Stenflo}}},
\batitle{{Resonance-line polarization. V - Quantum-mechanical interference
  between states of different total angular momentum}}.
\bjtitle{\aap}
\bvolume{84},
\bfpage{68}--\blpage{74}
(\byear{1980})
\end{barticle}
\endbibitem

\bibitem[\protect\citeauthoryear{{Stenflo}}{1982}]{stenflo-s82}
\begin{barticle}
\bauthor{\binits{J.O.} \bsnm{{Stenflo}}},
\batitle{{The Hanle effect and the diagnostics of turbulent magnetic fields in
  the solar atmosphere}}.
\bjtitle{\solphys}
\bvolume{80},
\bfpage{209}--\blpage{226}
(\byear{1982})
\end{barticle}
\endbibitem

\bibitem[\protect\citeauthoryear{{Stenflo}}{1985}]{stenflo-lest85}
\begin{barticle}
\bauthor{\binits{J.O.} \bsnm{{Stenflo}}},
\batitle{{LEST - A large international solar telescope for the 1990s}}.
\bjtitle{Vistas in Astronomy}
\bvolume{28},
\bfpage{571}--\blpage{576}
(\byear{1985})
\end{barticle}
\endbibitem

\bibitem[\protect\citeauthoryear{{Stenflo}}{1987}]{stenflo-s87}
\begin{barticle}
\bauthor{\binits{J.O.} \bsnm{{Stenflo}}},
\batitle{{Observational constraints on a 'hidden', turbulent magnetic field of
  the sun}}.
\bjtitle{\solphys}
\bvolume{114},
\bfpage{1}--\blpage{19}
(\byear{1987})
\end{barticle}
\endbibitem

\bibitem[\protect\citeauthoryear{{Stenflo}}{1993}]{stenflo-s93debate}
\begin{bchapter}
\bauthor{\binits{J.O.} \bsnm{{Stenflo}}},
\bctitle{{Strong and Weak Magnetic Fields: Nature of the Small-scale Flux
  Elements}},
in \bbtitle{IAU Colloq. 141: The Magnetic and Velocity Fields of Solar Active
  Regions},
ed. by \beditor{\binits{H.} \bsnm{{Zirin}}},
\beditor{\binits{G.} \bsnm{{Ai}}},
\beditor{\binits{H.} \bsnm{{Wang}}}
\bsertitle{Astronomical Society of the Pacific Conference Series} 
\bvolume{46},
\bfpage{205}--\blpage{214}
(\byear{1993})
\end{bchapter}
\endbibitem

\bibitem[\protect\citeauthoryear{{Stenflo}}{1994}]{stenflo-book94}
\begin{bbook}
\bauthor{\binits{J.O.} \bsnm{{Stenflo}}},
\bbtitle{{Solar Magnetic Fields --- Polarized Radiation Diagnostics.}}
(\bpublisher{Kluwer}, \blocation{Dordrecht}, \byear{1994})
\end{bbook}
\endbibitem

\bibitem[\protect\citeauthoryear{{Stenflo}}{2012}]{stenflo-s12aa2}
\begin{barticle}
\bauthor{\binits{J.O.} \bsnm{{Stenflo}}},
\batitle{{Basal magnetic flux and the local solar dynamo}}.
\bjtitle{\aap}
\bvolume{547},
\bfpage{93}
(\byear{2012})
\end{barticle}
\endbibitem

\bibitem[\protect\citeauthoryear{{Stenflo}}{2013a}]{stenflo-s13aa1}
\begin{barticle}
\bauthor{\binits{J.O.} \bsnm{{Stenflo}}},
\batitle{{Horizontal or vertical magnetic fields on the quiet Sun. Angular
  distributions and their height variations}}.
\bjtitle{\aap}
\bvolume{555},
\bfpage{132}
(\byear{2013}a)
\end{barticle}
\endbibitem

\bibitem[\protect\citeauthoryear{{Stenflo}}{2013b}]{stenflo-s13aarv}
\begin{barticle}
\bauthor{\binits{J.O.} \bsnm{{Stenflo}}},
\batitle{{Solar magnetic fields as revealed by Stokes polarimetry}}.
\bjtitle{\aapr}
\bvolume{21},
\bfpage{66}
(\byear{2013}b)
\end{barticle}
\endbibitem

\bibitem[\protect\citeauthoryear{{Stenflo}}{2014}]{stenflo-s14spw7}
\begin{bchapter}
\bauthor{\binits{J.O.} \bsnm{{Stenflo}}},
\bctitle{{Nature of Quiet-sun Magnetic Fields}},
in \bbtitle{Solar Polarization 7},
ed. by \beditor{\binits{K.N.} \bsnm{{Nagendra}}},
\beditor{\binits{J.O.} \bsnm{{Stenflo}}},
\beditor{\binits{Z.Q.} \bsnm{{Qu}}},
\beditor{\binits{M.} \bsnm{{Sampoorna}}}
\bsertitle{Astronomical Society of the Pacific Conference Series} 
\bvolume{489},
\bfpage{3}--\blpage{19}
 (\byear{2014})
\end{bchapter}
\endbibitem

\bibitem[\protect\citeauthoryear{{Stenflo} et~al.}{1980}]{stenflo-ik80}
\begin{barticle}
\bauthor{\binits{J.O.} \bsnm{{Stenflo}}},
\bauthor{\binits{D.} \bsnm{{Dravins}}},
\bauthor{\binits{N.} \bsnm{{Wihlborg}}},
\bauthor{\binits{A.} \bsnm{{Bruns}}},
\bauthor{\binits{V.K.} \bsnm{{Prokofev}}},
\bauthor{\binits{I.A.} \bsnm{{Zhitnik}}},
\bauthor{\binits{H.} \bsnm{{Biverot}}},
\bauthor{\binits{L.} \bsnm{{Stenmark}}},
\batitle{{Search for spectral line polarization in the solar vacuum
  ultraviolet}}.
\bjtitle{\solphys}
\bvolume{66},
\bfpage{13}--\blpage{19}
(\byear{1980})
\end{barticle}
\endbibitem

\bibitem[\protect\citeauthoryear{{Stenflo} et~al.}{1983}]{stenflo-setal83a}
\begin{barticle}
\bauthor{\binits{J.O.} \bsnm{{Stenflo}}},
\bauthor{\binits{D.} \bsnm{{Twerenbold}}},
\bauthor{\binits{J.W.} \bsnm{{Harvey}}},
\batitle{{Coherent scattering in the solar spectrum - Survey of linear
  polarization in the range 3165-4230 \AA}}.
\bjtitle{\aaps}
\bvolume{52},
\bfpage{161}--\blpage{180}
(\byear{1983a})
\end{barticle}
\endbibitem

\bibitem[\protect\citeauthoryear{{Stenflo} et~al.}{1983}]{stenflo-setal83b}
\begin{barticle}
\bauthor{\binits{J.O.} \bsnm{{Stenflo}}},
\bauthor{\binits{D.} \bsnm{{Twerenbold}}},
\bauthor{\binits{J.W.} \bsnm{{Harvey}}},
\bauthor{\binits{J.W.} \bsnm{{Brault}}},
\batitle{{Coherent scattering in the solar spectrum - Survey of linear
  polarization in the range 4200-9950 \AA}}.
\bjtitle{\aaps}
\bvolume{54},
\bfpage{505}--\blpage{514}
(\byear{1983b})
\end{barticle}
\endbibitem

\bibitem[\protect\citeauthoryear{{Stenflo} et~al.}{1984}]{stenflo-setal84}
\begin{barticle}
\bauthor{\binits{J.O.} \bsnm{{Stenflo}}},
\bauthor{\binits{S.} \bsnm{{Solanki}}},
\bauthor{\binits{J.W.} \bsnm{{Harvey}}},
\bauthor{\binits{J.W.} \bsnm{{Brault}}},
\batitle{{Diagnostics of solar magnetic fluxtubes using a Fourier transform
  spectrometer}}.
\bjtitle{\aap}
\bvolume{131},
\bfpage{333}--\blpage{346}
(\byear{1984})
\end{barticle}
\endbibitem

\bibitem[\protect\citeauthoryear{{Stepanov}}{1958}]{stenflo-stepanov58a}
\begin{barticle}
\bauthor{\binits{V.E.} \bsnm{{Stepanov}}},
\batitle{{The absorption coefficient of atoms in the case of reverse Zeeman
  effect for an arbitrary directed magnetic field}}.
\bjtitle{Izv. Krymsk. Astrofiz. Obs.}
\bvolume{18},
\bfpage{136}--\blpage{150}
(\byear{1958a})
\end{barticle}
\endbibitem

\bibitem[\protect\citeauthoryear{{Stepanov}}{1958}]{stenflo-stepanov58b}
\begin{barticle}
\bauthor{\binits{V.E.} \bsnm{{Stepanov}}},
\batitle{{On the theory of the formation of absorption lines in a magnetic
  field and the profile of Fe $\lambda$ 6173\,\AA\ in the solar sunspot
  spectrum}}.
\bjtitle{Izv. Krymsk. Astrofiz. Obs.}
\bvolume{19},
\bfpage{20}--\blpage{45}
(\byear{1958b})
\end{barticle}
\endbibitem

\bibitem[\protect\citeauthoryear{{Stepanov} and
  {Severny}}{1962}]{stenflo-stepseverny62}
\begin{barticle}
\bauthor{\binits{V.E.} \bsnm{{Stepanov}}},
\bauthor{\binits{A.B.} \bsnm{{Severny}}},
\batitle{{A photoelectric method for measurements of the magnitude and
  direction of the solar magnetic field}}.
\bjtitle{Izv. Krymsk. Astrofiz. Obs.}
\bvolume{28},
\bfpage{166}--\blpage{193}
(\byear{1962})
\end{barticle}
\endbibitem

\bibitem[\protect\citeauthoryear{{Trujillo Bueno}
  et~al.}{2004}]{stenflo-trujetal04}
\begin{barticle}
\bauthor{\binits{J.} \bsnm{{Trujillo Bueno}}},
\bauthor{\binits{N.} \bsnm{{Shchukina}}},
\bauthor{\binits{A.} \bsnm{{Asensio Ramos}}},
\batitle{{A substantial amount of hidden magnetic energy in the quiet Sun}}.
\bjtitle{\nat}
\bvolume{430},
\bfpage{326}--\blpage{329}
(\byear{2004})
\end{barticle}
\endbibitem

\bibitem[\protect\citeauthoryear{{Unno}}{1956}]{stenflo-unno56}
\begin{barticle}
\bauthor{\binits{W.} \bsnm{{Unno}}},
\batitle{{Line Formation of a Normal Zeeman Triplet}}.
\bjtitle{\pasj}
\bvolume{8},
\bfpage{108}
(\byear{1956})
\end{barticle}
\endbibitem

\bibitem[\protect\citeauthoryear{{Vrabec}}{1971}]{stenflo-vrabec71}
\begin{bchapter}
\bauthor{\binits{D.} \bsnm{{Vrabec}}},
\bctitle{{Magnetic Fields Spectroheliograms from the San Fernando
  Observatory}},
in \bbtitle{Solar Magnetic Fields},
ed. by \beditor{\binits{R.} \bsnm{{Howard}}},
\bsertitle{IAU Symposium} 
\bvolume{43},
\bfpage{329}--\blpage{339}
(\byear{1971})
\end{bchapter}
\endbibitem

\bibitem[\protect\citeauthoryear{{Zeeman}}{1897}]{stenflo-zeeman1897}
\begin{barticle}
\bauthor{\binits{P.} \bsnm{{Zeeman}}},
\batitle{{On the influence of Magnetism on the Nature of the Light emitted by a
  Substance}}.
\bjtitle{Phil. Mag.}
\bvolume{43},
\bfpage{226}
(\byear{1897})
\end{barticle}
\endbibitem

\bibitem[\protect\citeauthoryear{{Zirin}}{1993}]{stenflo-z93debate}
\begin{bchapter}
\bauthor{\binits{H.} \bsnm{{Zirin}}},
\bctitle{{The Interaction of Weak and Strong Magnetic Fields on the Sun
  (invited)}},
in \bbtitle{IAU Colloq. 141: The Magnetic and Velocity Fields of Solar Active
  Regions},
ed. by \beditor{\binits{H.} \bsnm{{Zirin}}},
\beditor{\binits{G.} \bsnm{{Ai}}},
\beditor{\binits{H.} \bsnm{{Wang}}}
\bsertitle{Astronomical Society of the Pacific Conference Series} 
\bvolume{46},
\bfpage{215}--\blpage{221}
(\byear{1993})
\end{bchapter}
\endbibitem

\end{thebibliography}
\end{document}